\documentclass[apj,onecolumn]{openjournal}
\usepackage{longtable}
\usepackage[hidelinks]{hyperref}
\usepackage{booktabs}
\usepackage{amsmath}
\usepackage{graphicx}
\usepackage{array}
\usepackage{tabularx}
\usepackage{makecell}
\usepackage{float}

\begin{document}

\title{Unveiling the Drivers of the Baryon Cycles with Interpretable Multi-step Machine Learning and Simulations}

\author{M. S. Khanom}
\author{B. W. Keller}
\author{J. I. Saavedra}

\affiliation{Department of Physics and Materials Science, University of Memphis,\\
3720 Alumni Avenue, Memphis, TN 38152, USA}

\begin{abstract}
We present a new approach for understanding how galaxies lose or retain baryons by utilizing a pipeline of two machine learning methods applied the IllustrisTNG100 simulation. We employed a Random Forest Regressor and Explainable Boosting Machine (EBM) model to connect the z=0 retained baryon fraction of $\approx10^5$ simulated galaxies to their properties. We employed Random Forest models to filter and used the five most significant properties to train an EBM. Interaction functions identified by the EBM highlight the relationship between baryon fraction and three different galactic mass measurements, the location of the rotation curve peak, and the velocity dispersion. This interpretable machine learning-based approach provides a promising pathway for understanding the baryon cycle in galaxies. 

\end{abstract}

\keywords{Baryonic matter --- Random Forest--- Explainable Boosting Machine--- IllustrisTNG100}

\section{Introduction} \label{sec:intro}
Baryons represent a small fraction of the cosmic energy density. \cite{Planck2016} reveals that baryons constitute just 5\% of the total energy density of the universe. The predicted amount of baryons is larger than the observed amount in galaxies and clusters, which has been dubbed the "missing baryon" problem (\cite{persic1992}). 

This missing matter is hypothesized to reside between galaxies in the warm-hot intergalactic medium (WHIM) or within the galaxy's diffuse circumgalactic medium (CGM). Recent studies utilizing the Chandra X-ray Observatory and the Hubble Space Telescope have presented some evidence for this missing component(e.g., \cite{Nicastro2018}).

\cite{Nicastro2018} identified two highly ionized oxygen (O VII) absorbers in the X-ray spectrum of a background quasar with a redshift above 0.4, concluding that the missing baryons are in the WHIM.  \cite{Gupta2012} analyzed Chandra X-ray observations, finding O VII and O VIII absorption lines in the Milky Way's CGM at about $10^6$ K. They found that the CGM extends over 100 kpc and contains over \(10^{10} \ M_{\odot}\) of baryons, indicating that the missing mass of the galaxy is in this warm-hot gas phase. \cite{Werk2014} the Cosmic Origins Spectrograph Halos (COS-Halos) survey L* galaxies have less than 6\% of their baryons in the extended warm CGM. In contrast, \cite{Li2018} reported that their fiducial galaxy has around 29\% of its baryons in the hot phase, with the extended hot halo mass making up about 45\% of the stellar mass within $R_{200}$. \cite{Nicastro2016} present a model for the distribution of hot gas in the Milky Way by analyzing O VII absorption in both the disk and halo. They suggest that the mass of this hot gas could explain the missing baryons within $R_{200}$. \cite{kovacs2019} applied a unique stacking technique to investigate the hot phases of the WHIM, revealing that the missing baryons are likely found as tenuous hot gas in the WHIM. Planck's Sunyaev–Zel'dovich (SZ) studies suggest that low metallicity hot gas could explain the missing baryons, with significant contributions from the extended CGM (\cite{Bregman2018}). Although observational studies have advanced our understanding of CGM and WHIM, they are limited by sensitivity and line-of-sight constraints. To interpret these observations, large-scale cosmological hydrodynamical simulations have been employed to study the distribution and evolution of baryons across cosmic time.

Recently, \cite{grauer2023} employed the IllustrisTNG50 simulation to estimate the X-ray optical depth($\tau$) of the intergalactic medium (IGM) at high z and compared it with gamma-ray bursts (GRB) afterglow observations. Their main findings include that without considering ionization, $\tau$ is overestimated by a factor 6, particularly at high z, where it approaches 0.9. After ionization, metals dominate X-ray observation ($>60\%$). The simulated IGM opacity ($\tau = 0.15 \pm 0.07 \text{ at } z = 10$) aligns with observed values ($\approx 0.4$) when accounting for residual host contributions and low metallicity. This suggests that the IGM, depending on its metallicity and ionization state, could account for a significant fraction of missing baryons.
\cite{Davies2020} also explored the relationship between the CGM and galaxy evolution using EAGLE and IllustrisTNG. They showed that low-mass halos ($M_{200} = 10^{11} M_{\odot}$) in TNG are more gas-rich ($f_{\text{CGM}}=0.55$) compared to EAGLE ($f_{\text{CGM}}<0.2$). Their study found strong correlations between $f_{\text{CGM}}$ and black hole mass, star formation rates, and galaxy morphology, driven by variations in feedback energy. \cite{Crain2007} showed that the baryon fraction within the virial radius of simulated haloes in the $\Lambda$CDM cosmology is typically 90\% of the cosmic average, with a 6\% rms scatter that remains consistent regardless of redshift. While simulations provide predictions of baryonic processes, their high dimensionality, non-linear interactions, and complex subgrid physics make it challenging to infer relationships between galaxy properties and the underlying physics. To address this, ML techniques have recently been incorporated to uncover patterns in simulation outputs and bridge the gap between simulations and observations.

Machine learning has recently become a powerful tool to infer relationships between features of simulated and observed galaxies. To model the relationship between baryonic properties and dark matter halos in the EAGLE simulations, \cite{Lovell2022} employ a tree-based learning technique known as Extremely Randomized Trees. \cite{machado2021shaping} used XGBOOST to model the complex relationship between gas shapes, dark matter, and baryonic density profiles in dark matter halos using IllustrisTNG hydrodynamical cosmological simulations. \cite{von2022inferring} used these same simulations to train supervised machine learning models in predicting dark matter halo properties in galaxies using optical and near-infrared imaging, as well as spectroscopy. \cite{Delgado2023} investigated the influence of feedback on matter clustering by training a random forest regressor on diverse feedback parameters and halo properties in the CAMELS simulations. \cite{Hausen2023} employed an EBM model to analyze data from the Cosmic Reionization on Computers (CROC) simulations, focusing on how dark matter halo properties  influence galaxies' star formation rate and stellar mass.

Observations and simulations propose various often conflicting locations for the missing baryons, including the Ly$\alpha$ forest, WHIM, and CGM. However, the distribution and physical state of baryons in the universe are still unresolved. At fixed halo mass, baryon fractions vary due to differences in internal processes such as feedback efficiency, gas inflows and outflows, formation history, environment, and structural properties all of which influence how effectively a galaxy retains or loses its baryonic content. The aim of this work is to improve our understanding of what physical mechanism sets the baryon fraction in simulated galaxies from IllustrisTNG100 using two machine learning models: EBMs and Random Forests.
We employed an EBM model to show the quantitative connection between galaxy properties and baryon fraction. As a pre-processing step, we used a Random Forest Regressor to select the important features in forecasting the baryon fraction.

EBMs are Generalized Additive Models (\cite{Hastie1986}) with automatic interaction terms. EBMs stand out because they provide clear explanations, in contrast to the complex and opaque nature of black-box models such as neural networks. Feature functions of a single variable or interaction functions of two variables in EBMs capture the dependencies of a target quantity (baryon fraction in our case) on each input parameter or pair of parameters. An EBM model is trained to fit these functions using a multivariate dataset. 

 EBM models are often described as interpretable since the magnitudes of the univariate functions ($f_{i}$) and bivariate functions ($f_{ij}$) directly indicate the relative importance of parameter ($\mu$) in producing the target quantity. If a certain parameter is not relevant, the EBM will find  $f_{i}= 0$ to obtain the desired quantity. The definition of the EBM can be found in Section 2.6.1.

 The layout of this work is structured as follows: Section 2 details the IllustrisTNG-100 simulation dataset, the preprocessing steps, and the machine learning techniques utilized to predict the retained baryon fraction. This section also discusses the performance metrics for evaluating prediction accuracy and details the features, target, and training procedures. Section 3 presents our findings, including the model's performance, feature importance analysis, univariate and bivariate interaction functions in predicting the baryon fraction of simulated galaxies. Section 4 provides a detailed discussion, including a comprehensive comparison with previous studies and future perspectives of our work. Section 5 concludes with a summary.

\section{Data and Methods} \label{sec:style}
\subsection{Why we need Multi-step Interpretable Machine Learning?}
Analyzing over 50 features in the dataset poses a challenge, even with an interpretable model. Manually selecting variables risks overlooking important features or adding bias due to our preconceptions, so we employed the permutation feature importance method with Random Forest to identify the most significant features. However, while permutation importance highlights which features are important, it does not reveal why they matter. To address this, we incorporated the EBM model to better understand the relationship between the target variable and the input features.

\subsection{The TNG 100 Simulations}
In this paper, we utilize the TNG100-1 simulation from the IllustrisTNG project (\cite{pillepich2018a,pillepich2018b, Nelson2018,Springel2018,marinacci2018,naiman2018,nelson2019}), a series of cosmological gravo-magnetohydrodynamical simulations. These simulations employ the Arepo moving-mesh code (\cite{Springel2010}) to follow the evolution of dark matter, gas, stars, and supermassive black holes. IllustrisTNG advances beyond its predecessor, the original Illustris simulation (\cite{Genel2014,Sijacki2015,Vogelsberger2014}) by incorporating an enhanced model of galactic physics (\cite{Pillepich2018,Weinberger2017}) and by integrating magnetic fields. 

The TNG100-1 simulation, which we use for this study, evolves a cube with a comoving side length of 100.7 Mpc. We only use $z=0$.

The TNG simulation uses specific cosmological parameters from the Planck 2015 data release \cite{Planck2016}. These parameters are $\Omega_\Lambda = 0.6911$, $\Omega_m = 0.3089$, $\Omega_b = 0.0486$, and $H_0 = 67.74 \, \text{km} \, \text{s}^{-1} \, \text{Mpc}^{-1}$. The resolution parameters of the simulation, including particle counts, softening lengths, and mass resolutions, are summarized in Table~\ref{Resolution_parameters}.

\begin{table}[h!]
    \centering
    \caption{Simulation Resolution Parameters}
    \begin{tabular}{|l|l|}
        \hline
        \textbf{Parameter} & \textbf{Value} \\ \hline
        Softening Length (Dark Matter \& Star Particles) & 0.74 kpc \\ \hline
        Softening Length (Gas Cells) & 0.185 kpc \\ \hline
        Mass Resolution (Dark Matter Particles) & $7.5\times10^6 \, M_\odot$ \\ \hline
        Mean Mass Resolution (Baryon Particles) & $1.4\times10^6 \, M_\odot$ \\ \hline
    \end{tabular}
    \label{Resolution_parameters}
\end{table}

The baryonic TNG runs include radiative cooling influenced by a redshift-dependent ionizing background, self-shielding corrections, and stochastic star formation within dense interstellar medium (ISM) gas. They also model ISM pressurization from unresolved supernovae using an effective equation of state (\cite{springel2003}) and simulate stellar population evolution with associated chemical enrichment and gas recycling from supernovae and stellar winds. Stellar feedback drives galactic-scale outflows through an energy-driven, kinetic wind mechanism (\cite{nelson2019}). 

The IllustrisTNG model incorporates AGN feedback operating in two distinct modes (\cite{weinberger2016}). The thermal feedback mode, active during high black hole accretion rates (\(\dot{M}_{\text{BH}} \geq 0.1 \dot{M}_{\text{Edd}}\)), injects thermal energy. At lower accretion rates, the kinetic feedback mode drives directional outflows. This approach addresses shortcomings in the original Illustris simulation, where feedback overly expelled gas in galaxy groups and clusters (\cite{Genel2014}). 

 Halos are identified through the ``friends-of-friends'' (FOF) algorithm, which forms clusters by connecting dark matter particles that are $< 0.2$ times the average particle spacing. Subhalos are recognized using the SUBFIND algorithm, which requires each subhalo to comprise a minimum of 20 gravitationally bound dark matter particles. We used the z = 0 snapshot, which represents present-day galaxy structures. This snapshot contains 1,048,574 subhalos in the object catalog. For our analysis, we chose 107,867 subhalos, which we will refer to as galaxies. 

\subsection{Data Preprocessing}
Data preprocessing is a crucial stage for machine learning.
We considered the most bound subhalo in each FOF parent halo with $M_{\text{gas}} > 0$ and $M_{\text{star}} > 0$.  We only consider halos resolved with $\geq 1000$ DM particles, giving $M_{\text{200}} \geq 4.5 \times 10^9$, where \( M_{\text{200}} \) refers to the mass of the FOF group (Group\_M\_Crit200).  To avoid baryonic substructure mis-identified as halos, we select only groups with total baryon fractions below the cosmological average of 0.16. At the subhalo level, we included only those with $SubhaloFlag == 1$, excluding halos that SUBFIND determines to be spurious.

To prepare the data for training the model, we applied $\log_{10}$ transformations to all features in training and testing datasets because these features span many orders of magnitude. Many of our datasets contain zero values. For these, we replaced the zero values of each feature with a small positive number $\epsilon = 10^{-4}\min(\mu_i)$. 

After processing, our dataset comprised 89 features across 107,867 galaxies. To address multicollinearity, we employed the Pearson correlation method, identifying pairs of features with strong correlations (correlation coeffcients $> 0.75$). Then we applied the Variance Inflation Factor(VIF) method to mitigate the multicollinearity issues among the features (\cite{salmeron2020}). We also used stepwise linear regression to select features for training the model. We dropped 23 features due to strong multicollinearity and leakage issues, which could negatively impact the model performance by introducing redundancy. By removing these highly correlated features, we simplified the feature set and reduced the potential for multicollinearity. The correlation matrices in Appendix B display 27 features that were evaluated for redundancy based on pairwise correlations. From this set, 23 features were removed using the VIF and regression filtering. A list of removed features is provided in Appendix B.

 \subsection{Machine Learning models}
 Figure~\ref{fig:pipeline} shows our data processing pipeline. We first select 66 features from IllustrisTNG-100 for 107,867 simulated galaxies. A Random Forest Regressor was trained on 75\% of the data, leaving 25\% on the test set. A permutation importance method on the trained Random Forest selected the five most important features, which were then used to train an EBM model. The EBM model allowed us to explore how these feature interactions influence baryon retention.
\begin{figure}
  \centering
  \includegraphics[width=\linewidth]{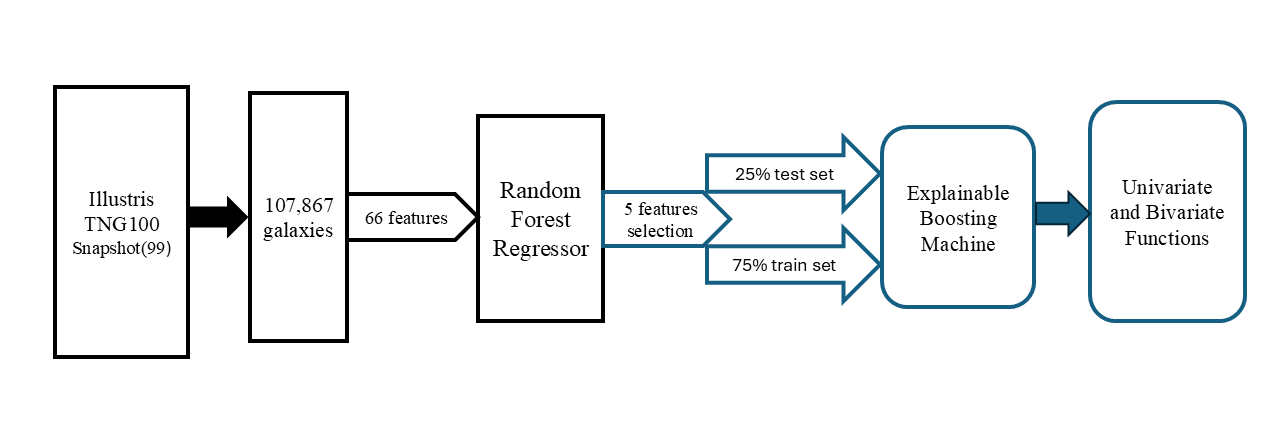}
    \caption{\textit{The processing pipeline is used in our work. Initially, we extracted data from the IllustrisTNG simulation, including 107,867 simulated galaxies and 66 features. Then, a Random Forest model was used to identify the top 5 important features. An EBM was then trained on these features using a 75\%/25\% train-test split to analyze their relationships with the target variable. Finally, the EBM results explored univariate and bivariate functions, providing insights into galaxy properties.}}
    \label{pipeline}
 \end{figure}

\subsubsection{Random Forest Model}
We employed the random forest regressor from the Scikit-Learn package (\cite{pedregosa2011}). This model comprises several decision trees, each a separate regression model. A decision tree is a predictive model that divides data into branches at decision points, forming a tree structure. These trees are trained on discrete random subsets of the training set (\cite{Breiman2001}). The average of the predictions made by these trees is the random forest's total output, which helps reduce overfitting, a typical issue with single-decision trees. Many cosmological investigations (\cite{Cohn2020,Lucie-Smith2018,Nadler2018,gensior2024}) have demonstrated the effectiveness of random forests in handling complex and high-dimensional data. 

\subsection{Hyperparameter Tuning for Random Forest Model}
The hyperparameters we used were tuned during sklearn's GridSearchCV hyperparameter selection tool. We examined five hyperparameters that early experiments identified as promising. These were n\_estimators, number of decision trees; max\_depth, the depth these trees are allowed to reach; min\_samples\_split, which determines how many samples are required to add a layer to the tree; min\_samples\_leaf, which determines the number of samples in the smallest leaf node; and max\_samples, the fraction of the full dataset used per tree.

The range of values we test and the optimal values are shown in Table~\ref{hyperparameter_values}.

\begin{table}
\caption{Hyperparameter Values Tested and Optimal Values Found During Tuning for the Random Forest Model}
\begin{center}
\begin{tabular}{| c | c |c|}
\hline
Hyperparameter & Values Tested & Optimal values \\
\hline
\hline
\textbf{n\_estimators} & [50,100, 200, 300, 400, 450, 500, 1000] & 300\\
\textbf{max\_depth} & [None, 10, 20, 30, 35, 40] & 30\\
\textbf{min\_samples\_split} & [2, 5, 10, 15, 20, 25] & 15\\
\textbf{min\_samples\_leaf} & [1, 2, 4, 6, 8] & 6\\
\textbf{max\_samples} & [0.5, 0.75, 0.9] & 0.75\\
\hline
\end{tabular}
\end{center}
\label{hyperparameter_values}
\end{table}

\subsection{Selecting Features using Permutation Importance on the Random Forest}
Feature importance is a technique to identify which features contribute to the prediction accuracy of machine learning models. The permutation importance \( I_{j} \) can be expressed as (\cite{Breiman2001}):
\[
I_{j} = S_{\text{ref}} - \frac{1}{N} \sum_{r=1}^{N} S_r(\mu_j)
\]
Where \( S_{\text{ref}} \) is the model's original performance score before permutation, \( S_r(\mu_j) \) is the model's performance score when feature \( \mu_j \) has been randomly shuffled in repetition \( r \) and \( N \) is the number of repetitions for shuffling and scoring (100 in our case). 

We used this approach to identify and rank the most significant features for predicting the retained baryon fractions using the Random Forest Model, which are then fed to the EBM to determine how these features impact baryon fraction.

\subsection{Explainable Boosting Machine}
We use a type of  Generalized Additive Model (GAM)  known as Explainable Boosting Machines (EBMs). GAMs are defined as
\begin{equation}
Y(\vec{\mu}) = \beta + \sum_{i} f_{i}(\mu_{i})
\label{GAM_definition}
\end{equation}

Here, $\beta$ is the intercept, $Y(\vec{\mu})$ is the prediction given feature $\vec{\mu}$, and $f_i$ are learned (non-parametric) functions. These functions affect each input feature $\mu_i$ individually (\cite{Hastie1990}). GAMs are less flexible than many other machine learning models because of their limited capacity to capture feature interactions. While both GAMs and EBMs offer a representation of how target variables $y$ are related to parameters $\vec{\mu}$. EBMs advance this structure by adding both univariate ($f_i(\mu_i)$) and bivariate functions $f_{ij}(\mu_i,\mu_j))$ to represent better the interactions between parameter pairs and their combined impact on the target variable $Y$.

\begin{equation}
Y(\vec{\mu}) = \ \beta + \ \sum_{}^{}{f_{i}\left( \mu_{i} \right)} + \sum_{}^{}{f_{ij}{\ (\mu}_{i}, \mu_{j})}
\end{equation}

These functions are learned using gradient boosting to learn these functions (\cite{friedman2001,breiman2001a}). In boosting, many decision trees are built sequentially to develop a robust predictive model, with each tree focusing on correcting the errors of the previous ones, while in a Random Forest, decision trees are built independently and in parallel, with each tree contributing equally to the final prediction. This technique reduces overfitting, leading to stable predictions. 

\subsection{Features and Targets}
Our primary focus in this study is to identify the galaxy and halo properties most strongly associated with the baryon fraction within galaxies. The baryon fraction is the target variable of our multi-step machine learning pipeline:

\[f_{\text{Bar}} = \frac{M_{\text{star}} + M_{\text{gas}}}{M_{\text{200}}}\]

Here, \( M_{\text{star}} \) is the mass of the stars of the subhalo, 
\( M_{\text{gas}}\) is the mass of the gas of the subhalo, and 
\( M_{\text{200}}\) is the total mass of the corresponding FOF group (as identified by FOF). 

In our case, we initially considered a comprehensive set of 66 features associated with baryonic matter within the halos and sub-haloes, as outlined in the IllustrisTNG100 halo catalog. We subsequently used a random forest to pinpoint the 5 features with the highest importance scores, which appear in the top 5 in all 20 independent training sessions. The top 5 features are shown in the Table~\ref{Most_significant_features} below:

\begin{table}[htbp]
\centering
\caption{The 5 most crucial features were identified by their highest importance scores using permutation feature importance from the Random Forest model. Below are these features, along with their definitions, units, and corresponding importance values.}

\begin{tabular}{| l | l | l | l |}
\hline
\textbf{Variable} & \textbf{Unit} & \textbf{Description} & \textbf{Importance}\\
\hline
$M_{\text{gas, MaxRad}}$ & $M_{\odot}$ & The gas mass within $R_{V_{\text{Max}}}$ & $1.799 \pm 0.00076$\\
\hline
$M_{\text{SFG}}$ & $M_{\odot}$ & Total mass of gas actively forming stars ($SFR >0$) & $0.263 \pm 0.00088$\\
\hline
$M_{200}$ & $M_{\odot}$ & The halo mass & $0.176 \pm 0.00048$\\
\hline
$R_{V_{\text{Max}}}$ & kpc & The radius of the rotation curve peak ($V_{\text{Max}}$) & $0.122 \pm 0.00016$ \\
\hline
$\sigma$ & km/s & The velocity dispersion of all particles in the galaxy & $0.030 \pm 0.0010$ \\
\hline
\end{tabular}
\label{Most_significant_features}
\end{table}

\subsection{Training and Test samples}
 
\subsubsection{Hyperparameter Tuning for EBM Model}
As with our Random Forest, we use GridSearchCV to obtain a tuned set of hyperparameters. 
Table~\ref{ebm_hyperparameters} displays the optimal hyperparameters. The hyperparameters we set are the optimization scheme's learning rate, number of bins for univariate and bivariate functions, smoothing rounds and other parameters mentioned in table 2 throughout the fitted domain.

\begin{table}
\centering
\caption{The optimal hyperparameters for the EBM model were identified using the GridSearchCV approach. All other remaining hyperparameters were kept at their default settings as specified in InterpretML version 0.6.2. }

    \begin{tabular}{ | l | l | l|}
 \hline
 \textbf{Hyperparameter}  & \textbf{Values Tested} & \textbf{Optimal Values} \\
 \hline
 \hline
 Binning & ``Quantile'' & ``Quantile'' \\
 \hline
 Maximum Bins for Univariate function & [256] & 256\\
 \hline
 Maximum Bins for Bivariate function & [32] & 32\\
 \hline
 Learning Rate & [0.1,0.2,0.3,0.4,0.5,0.6,0.7,0.8,0.9,1.0] & 0.6 \\
 \hline
 Outer Bags & [4,14,20] & 14\\
 \hline
 Inner Bags & [4,8,12] & 8 \\
 \hline
 Interactions & [20] & 20 \\
 \hline
 Minimum Samples Leaf & [2,10,15,20] & 20 \\
 \hline
 Smoothing Rounds & [200,500,1000,1500,2000,3000,5000] & 2000 \\
 \hline
\end{tabular}
\label{ebm_hyperparameters}
\end{table}

An important point to emphasize is that we employed the hold-out method, using an entirely unseen testing set to evaluate the model's performance. This testing set was excluded from the model's training process and was not involved in the cross-validation (CV) used for hyperparameter optimization. This ensures that the model's quality is evaluated on data that it has not been trained on. When evaluating a model on a testing set, the predictive performance is usually more conservative than the results obtained during training with CV (\cite{Hastie2009,Torgo2011}).

\subsection{Performance Metrics}
We employed well-established statistical metrics; the Coefficient of Determination ($R^2$ score) and Mean Absolute Error (MAE); to quantify the predicted values' accuracy. These metrics assess the model's performance by comparing the predicted values to the ground truth values from the test sample. 
The $R^2$ score, representing the proportion of the variance in the target values explained by the model (\cite{chicco2021}), ranges from 0 to 1, where one suggests that the model fits the data perfectly, capturing 100\% of the variance in the target variable. Conversely, a score close to 0 indicates a poor fit. 

The coefficient of determination $R^2$ is defined mathematically as 
\begin{equation}
R^2 = 1 - \sum_{\alpha}\ \frac{\left( Y_{\alpha} - {\widehat{Y}}_{\alpha} \right)^{2}}{\left( Y_{\alpha} - \Bar{Y}\right)^{2}}\
\end{equation}
Where
\(Y_{\alpha}\) represents the actual (observed) value for the \( \alpha\text{th} \) data point. \({\widehat{Y}}_{\alpha}\) represents the predicted value for the \( \alpha\text{th} \) data point and \( \Bar{Y}\) is the mean of the actual values.

MAE is the average magnitude of the prediction errors without considering their direction. Lower value of MAE signifies better model performance.
The MAE of any model calculated applied to the true values are:
\begin{equation}
\text{MAE} = \frac{1}{N} \sum_{\alpha=1}^{N} \left| \widehat{Y}_{\alpha} - Y_{\alpha} \right|
\end{equation}

where N is the number of data points.

\section{Results} \label{sec:Results}
\subsection{Model Performance}
After training the model, we apply performance metrics (discussed in 2.9) to the test set – better metrics indicate a better model (lower for error-based metrics like MAE, higher for goodness-of-fit metrics like $R^2$). 
\subsubsection{Performance Metrics for Random Forest}
Table~\ref{performance_metrics_RF} displays the results for performance metrics ($R^2$ and MAE) for Random Forest model discussed below, computed for both the training and test samples.

\begin{table}[h]
\centering
\caption{Performance Metrics for the Trained RF Model}
   \begin{tabular}{| l | l | l |}
 \hline
\textbf{Metrics} & \textbf{Training Set} & \textbf{Testing Set} \\
 \hline
 \hline
 Coefficient of Determination, $R^2$ & 0.94 & 0.897 \\
 \hline
 Mean Absolute Error (MAE) & 0.005 & 0.007 \\
 \hline
\end{tabular}
\label{performance_metrics_RF}
\end{table}
 The error metrics in the test set are within 10\% the training set, indicating minimal overfitting.  Since Random Forest was employed for feature selection, the minimal overfitting observed is considered acceptable and does not significantly affect the overall performance of the EBMs. 
 
\subsubsection{Performance Metrics for EBM}
Table~\ref{performance_metrics_EBM} displays the results for performance metrics ($R^2$ and MAE) for EBM discussed below, computed for both the training and test samples. 

\begin{table}[h]
\centering
\caption{Performance Metrics for the Trained EBM Model}
   \begin{tabular}{| l | l | l |}
 \hline
\textbf{Metrics} & \textbf{Training Set} & \textbf{Testing Set} \\
 \hline
 \hline
 Coefficient of Determination, $R^2$ & 0.867 & 0.866 \\
 \hline
 Mean Absolute Error (MAE) & 0.008 & 0.008 \\
 \hline
\end{tabular}
\label{performance_metrics_EBM}
\end{table}
No overfitting issues are observed in the EBM model as the performance metrics both for training and testing sets align well. 

\begin{figure}[h]
    \centering
    \includegraphics[width=\textwidth]{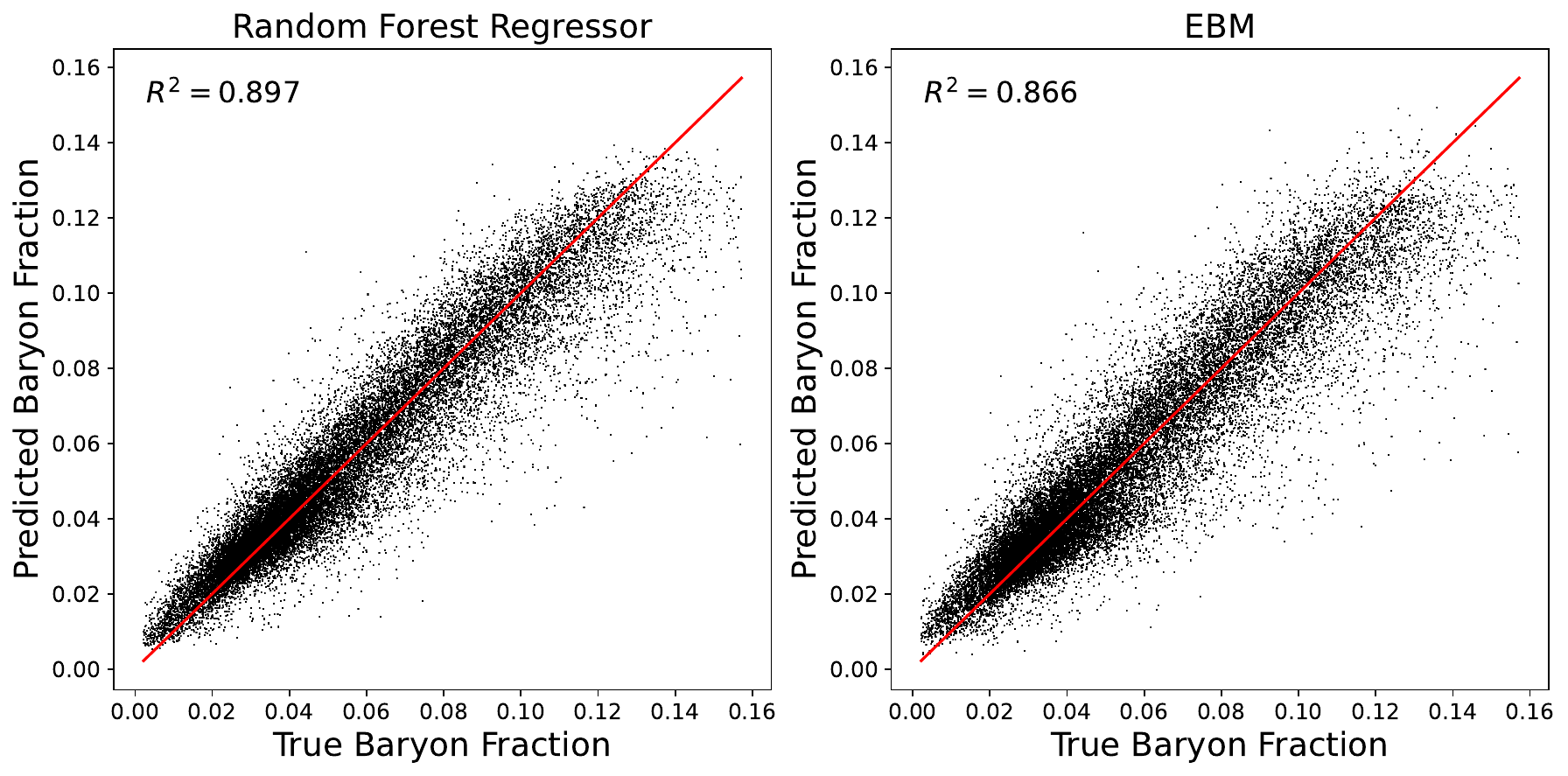}
    \caption{\textit{Accuracy of the two ML models at predicting $f_Bar$: the Random Forest Regressor (left) with 66 features and the EBM (right) with 5 features. The red line shows where predictions equal the true values, demonstrating a strong correlation with $R^2$ scores of 0.897 for the Random Forest model and 0.866 for the EBM.}}
    \label{scatterplots_RF_EBM}
 \end{figure}

 Figure~\ref{scatterplots_RF_EBM} displays our model's predictions alongside the true baryon fractions in the test set of galaxies both for Random Forest Regressor and EBM. The high ($R^2$ = 0.897) score of Random Forest model indicates a relatively strong fit. Reducing the number of features from 66 to 5 has only a small impact on our $R^2$ score with the EBM.
 
\subsection{Feature Importances}

Table~\ref{Most_significant_features} depicts the top 5 important features from our Random Forest Regressor model that targets the retained baryon fraction. The feature $M_{\text{gas, MaxRad}}$ (gas mass within the radius of $V_{Max}$), is the most important feature. $M_{\text{SFG}}$ (star forming gas mass) ranks as the second most important feature for predicting baryon fraction while $\sigma$ (velocity dispersion) noted as the least significant. $M_{200}$, and $R_{V_{\text{Max}}}$ contribute at the few-percent level.

 The uncertainty column is the standard deviations of the feature importance scores derived from 10 different subsets of the dataset, essentially applying a bootstrap to the permutation importance algorithm. 

\subsection{Targeting Baryon Fraction with an EBM Model}
\begin{figure}[h]
    \centering
    \includegraphics[width=1\textwidth]{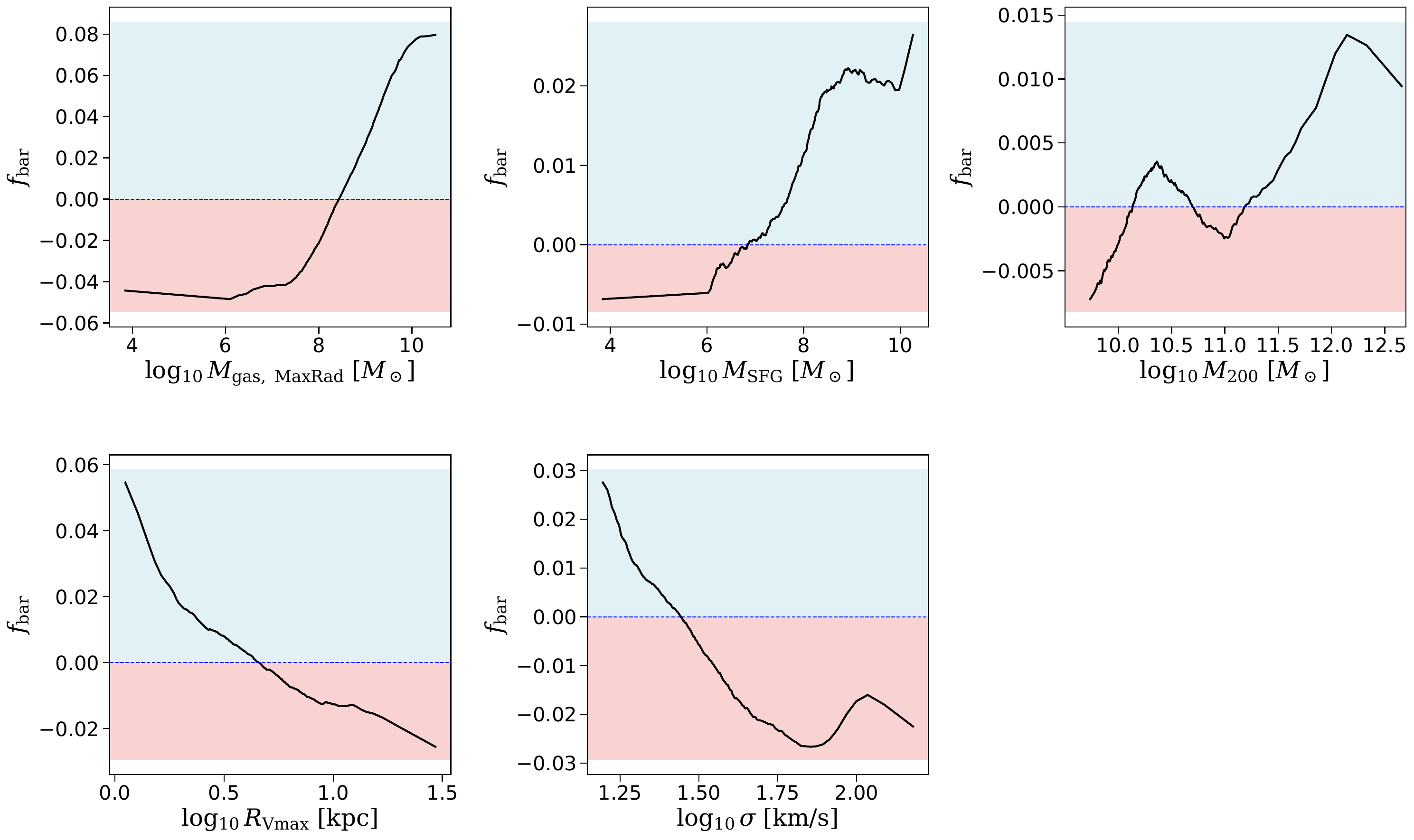}
    \caption{\textit{The univariate feature functions ($f_{i}$) for the EBM model trained to predict the baryon fractions in galaxies. From the left to right, the feature functions correspond to gas mass within the $R_{V_{Max}}$ ($M_{\text{gas, MaxRad}}$), star-forming gas mass($M_{SFG}$), halo mass ($M_{200}$),  radius at the maximum rotational velocity ($R_{V_{Max}}$), velocity dispersion ($\sigma$). Light blue areas above zero indicate regions where $f_{i}>0$, while light coral areas below zero indicate regions where $f_{i}<0$. Negative values of $f_{Bar}$ indicate a reduction in the predicted baryon fraction relative to the baseline ($\beta = 0.0542$), not a negative baryon fraction. The total prediction remains positive when all contributions, including $\beta$ are summed.}}
    \label{univariate_features_importance}
 \end{figure}
Figure~\ref{univariate_features_importance} displays the plots of the univariate feature functions $f_{i}$ for every feature. The functions show that the baryon fraction is increasing with  $M_{\text{gas, MaxRad}}$ and $M_{\text{SFG}}$, i.e., these univariate functions show an increasing trend with the baryon fraction. 

The baryon fraction increases with $M_{\text{gas, MaxRad}}$, indicating a positive correlation. This positive correlation indicates that galaxies with more gas mass within the radius of $V_{Max}$ retain more baryons.

The positive correlation between $M_{\text{SFG}}$ and the baryon fraction suggests that galaxies with more dense, cool gas in their disks have higher overall baryon fractions.

The baryon fraction decreases with increasing $R_{V_{Max}}$. This suggests that more compact galaxies, with deeper gravitational potential wells, are better at retaining baryons. Being centrally concentrated, bulge-dominated galaxies are effective at preventing ejection, resulting in higher baryon retention. 

\begin{figure}[h]
    \centering
    \includegraphics[width=1\textwidth]{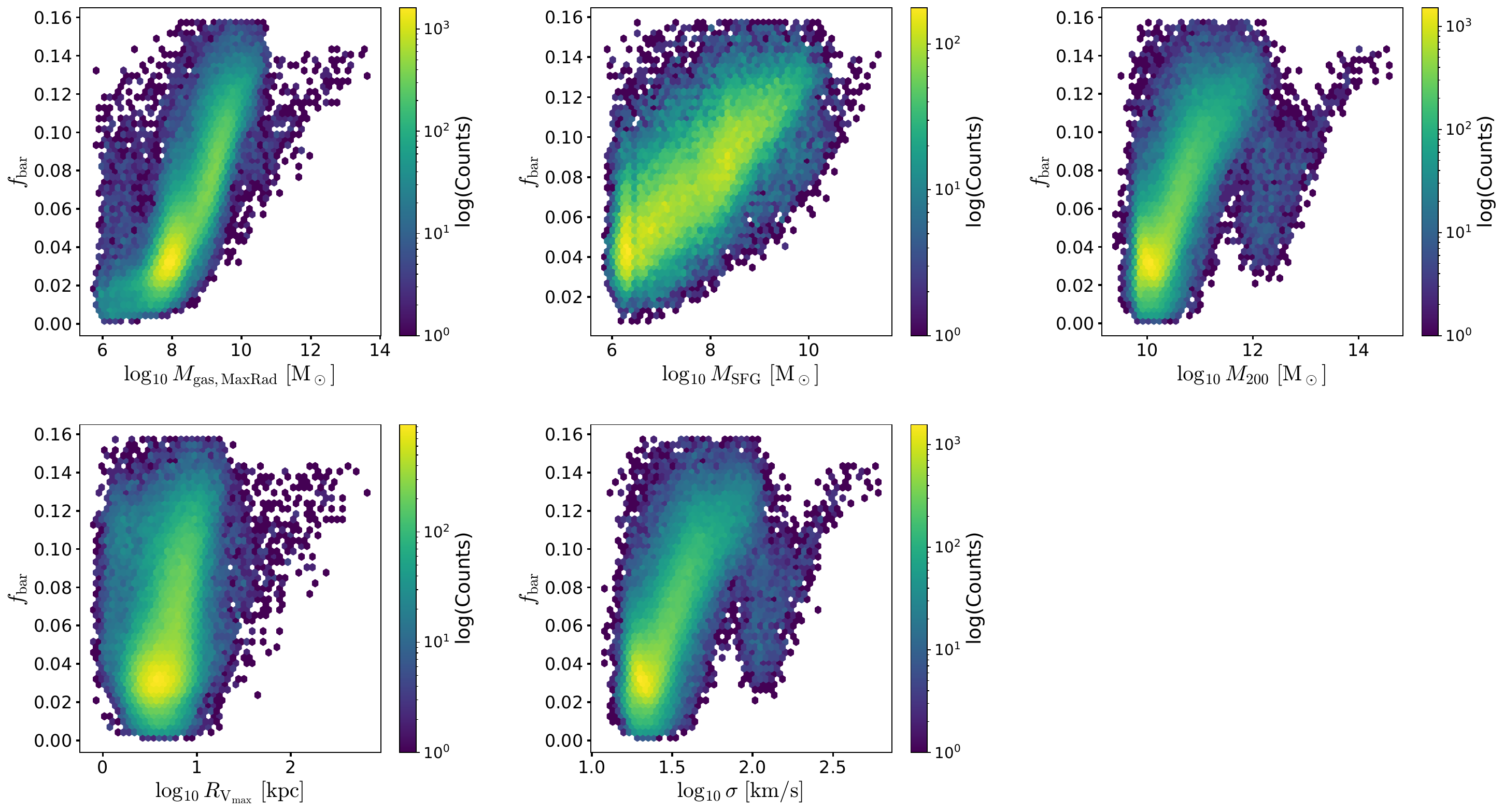}
    \caption{Direct relationships between baryon fraction and each of the five most important features identified by the Random Forest model, shown as raw scatterplots.}
    \label{fbar_five_features}
\end{figure}

Figure~\ref{fbar_five_features} displays the direct relationships between the retained baryon fraction and the five most important features as raw scatterplots. These plots show marginal correlations, examining each feature individually while other galaxy properties vary. Thus, they do not directly correspond to the EBM univariate feature functions shown in Figure~\ref{univariate_features_importance}. The raw scatterplots show overall trends with substantial scatter and correlations, while the EBM univariate functions reflect conditional additive contributions learned after accounting for correlations among all input features.  
\begin{figure}[h]
    \centering
    \includegraphics[width=\textwidth]{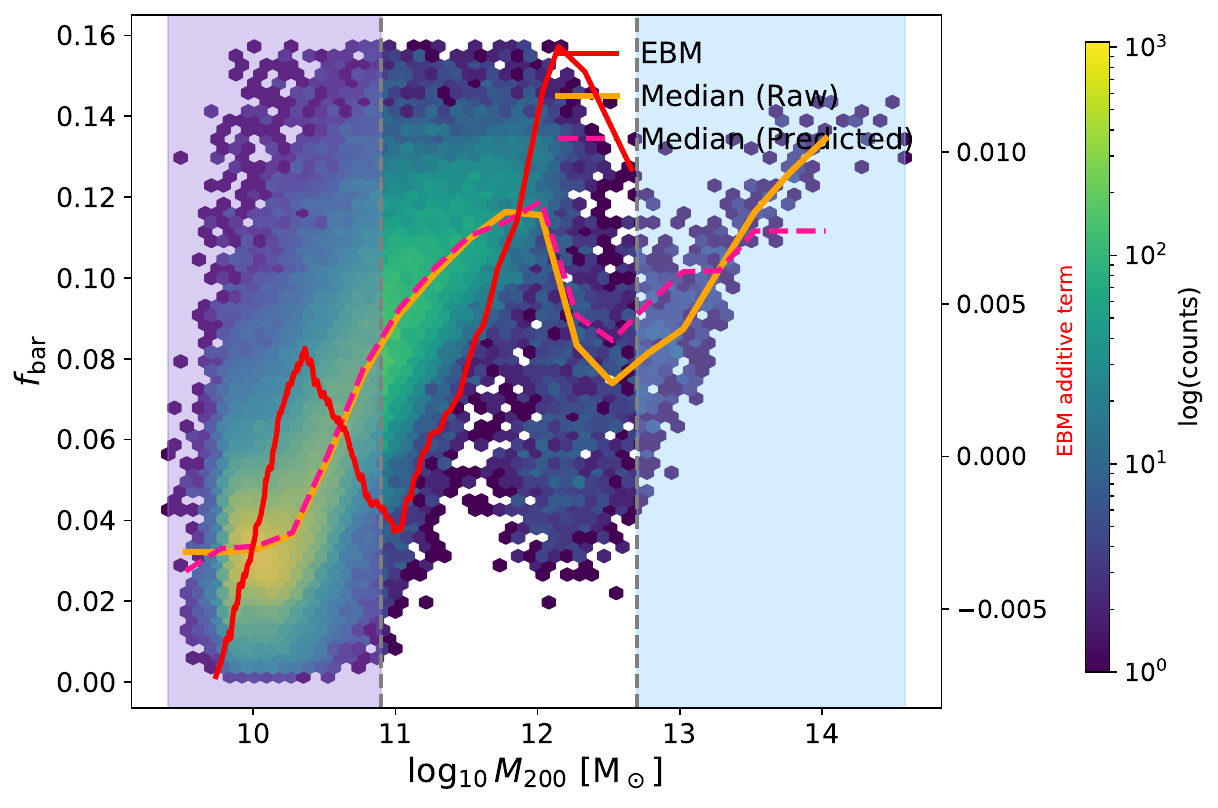}
    \caption{\textit{Retained baryon fraction $f_{bar}$  as a function of $M_{200}$. The hexbin background shows the raw galaxy distribution (log-scaled counts), and the red line shows the EBM univariate function (right y-axis), and the orange line indicates the median trend from binned data (left y-axis). The dashed hot pink curve shows the median of the full EBM prediction (left y-axis). The purple-shaded region marks the low-mass range not included in the \cite{wright2024}, while the skyblue-shaded region highlights the high-mass regime where the number of halos is very small (380).}}
    \label{scatterplot_M200}
 \end{figure}
 
Figure~\ref{scatterplot_M200} illustrates the distinction between the isolated EBM univariate contribution associated with $M_{200}$ and the full EBM prediction. Although the univariate EBM function for $M_{200}$ shows a trough near $\log_{10} M_{200}\sim 11$, the median of the full EBM prediction closely follows the median trend of the raw data throughout the full halo-mass range. This demonstrates that the apparent trough in $f_{(M_{200})}$ does not correspond to a physical suppression of the baryon fraction at that mass scale; rather, it reflects how the EBM distributes predictive influence across correlated variables such as $M_{200}$ and $\sigma$ rather than attributing all predictive power uniquely to $M_{200}$.

Notably, the relationship between $\sigma$ and $M_{200}$, shown in the Figure~\ref{Virial_relation}, reveals a tight, nearly one-to-one relationship, indicating that they are strongly correlated and largely encode the same information. The small scatter around the relation reflects residual differences between systems. We interpret $M_{200}$ and $\sigma$ act as proxies for one another.

The relationship between baryon fraction and halo mass shows a non-monotonic trend. The observed positive correlation at higher halo masses is what we would expect: that more massive halos, with their deeper gravitational potential wells, are better at retaining baryons. Figure 4, which uses raw data, shows a decline in baryon fraction in the $M_{200} \approx 10^{12}\text{--}10^{12.5}\,M_{\odot}$ range consistent with the findings of \cite{wright2024}. 
In our EBM analysis, the univariate curve for $M_{200}$ rises from $\log_{10} M_{200}\sim 11$ to a peak near $\log_{10} M_{200} \sim 12$ and then declines, which is consistent with the trend reported by \cite{wright2024}. At the highest masses, the EBM curve does not reproduce the full upturn seen in \cite{wright2024}., which we attribute to limited sample size and binning effects: the EBM model discretizes continuous features into bins using quantile (We used max\_bins=256 in our work). As a result, low-density regions at the extreme ends of the $M_{200}$ distribution (e.g., high-mass halos) may be under-resolved or entirely excluded from the learned shape functions. Nevertheless, our raw scatterplots (Figure 4) show good agreement with Wright et al. The decline from $\log_{10} M_{200} \sim 12$ to $12.5$ may be due to the onset of AGN-driven outflows that lower the baryon fraction. In the region $\log_{10} M_{200} \sim 11{-}12$ our results overlap  Wright et al. very well, while below this range the EBM curve more closely follows the average trend seen in our raw scatter plots(Figure 4).

Overall, as the velocity dispersion increases, the baryon fraction decreases. This suggests that galaxies with higher velocity dispersions tend to lose more baryons.   

\subsection{EBM Based Prediction of Baryon Fraction: Insights from Bivariate Features}
\begin{figure}[h]
    \centering
    \includegraphics[width=1\textwidth]{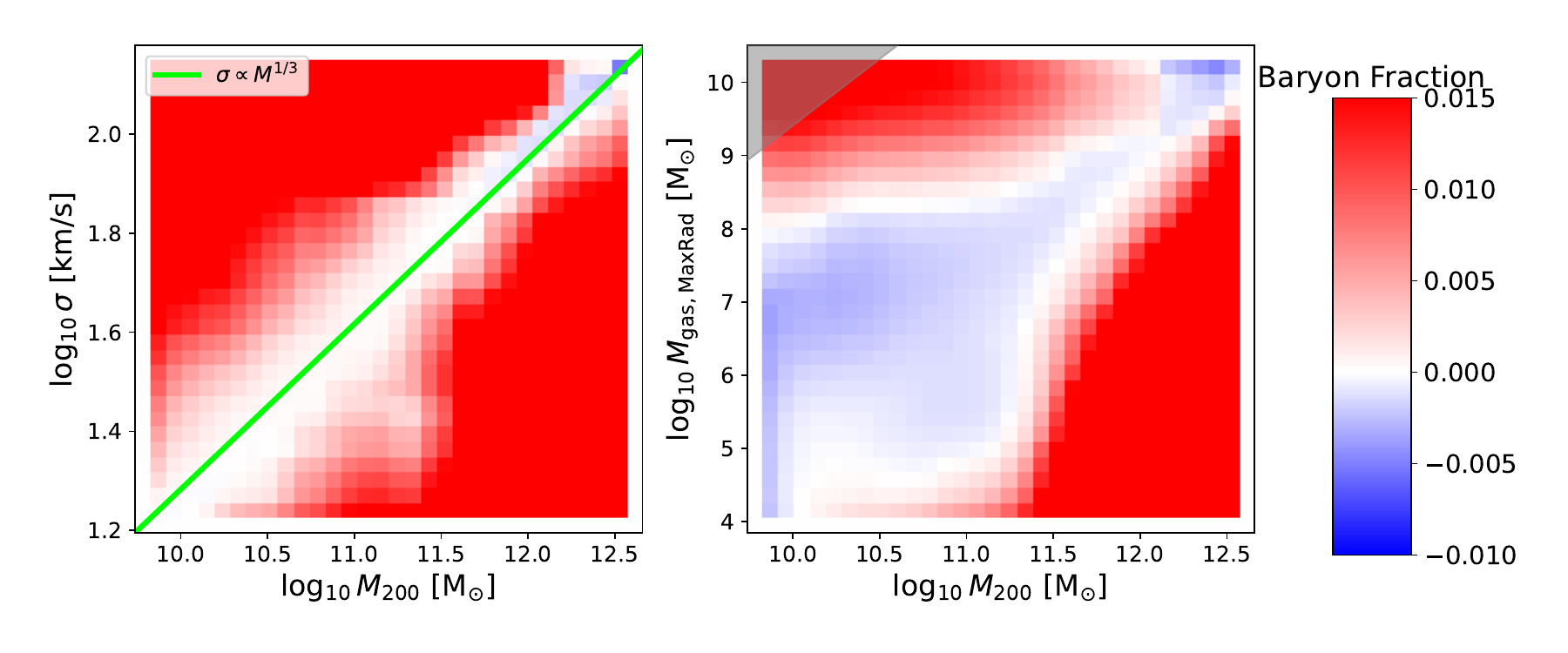}
    \caption{\textit{The bivariate interaction functions \( f_{ij} \) for the EBM model. Each panel shows the contribution of these interactions, with colors ranging between positive and negative values, normalized by the maximum norm of each function \( \|f\|_{\mathrm{max}} \). Red indicates areas where the feature interactions positively contribute to the baryon fraction, while dark blue represents regions with negative contributions. The most significant interaction occurs between \( M_{200} \) and \( \sigma \). Another notable interaction is between \( M_{\text{gas, MaxRad}} \) and \( M_{200} \). The remaining interaction functions are relatively weak, contributing only minor variations (on the order of \( \sim 0.01\approx0 \)) to \( f_{\mathrm{bar}} \), and are not shown here.}}

    \label{bivariate_features_importance}
 \end{figure}

The bivariate functions $f_{ij}$ learned for the EBM model are displayed in the heatmap shown in Figure~\ref{bivariate_features_importance}. The contributions of the interaction functions are lower than the univariate functions.

The interaction between $M_{200}$ and $\sigma$ reflects the proportional relationship in virial equilibrium, where $M_{200}\propto \sigma^{3}$. This heatmap reveals that galaxies that deviate from this expected relation, i.e., those outside virial equilibrium, retain more baryons. Furthermore, the interaction of $M_{\text{gas, MaxRad}}$ with $M_{200}$ shows that galaxies with low gas mass in the center have lower baryon fractions if they reside in low-mass halos, due to their shallow gravitational potential wells. In contrast, massive halos with low central gas mass tend to have higher baryon fractions. Galaxies with both high halo mass and gas-rich retain more of their baryons because of their deeper gravitational potential wells.The gray-shaded region is physically excluded, as it would imply that the gas mass within the central region exceeds the total expected baryonic mass of the halo, which is unphysical.

\subsection{Handling Highly and Physically Correlated Features: $M_{200}$ and $\sigma$}

\begin{figure}[h]
    \centering
    \includegraphics[width=0.8
    \textwidth]{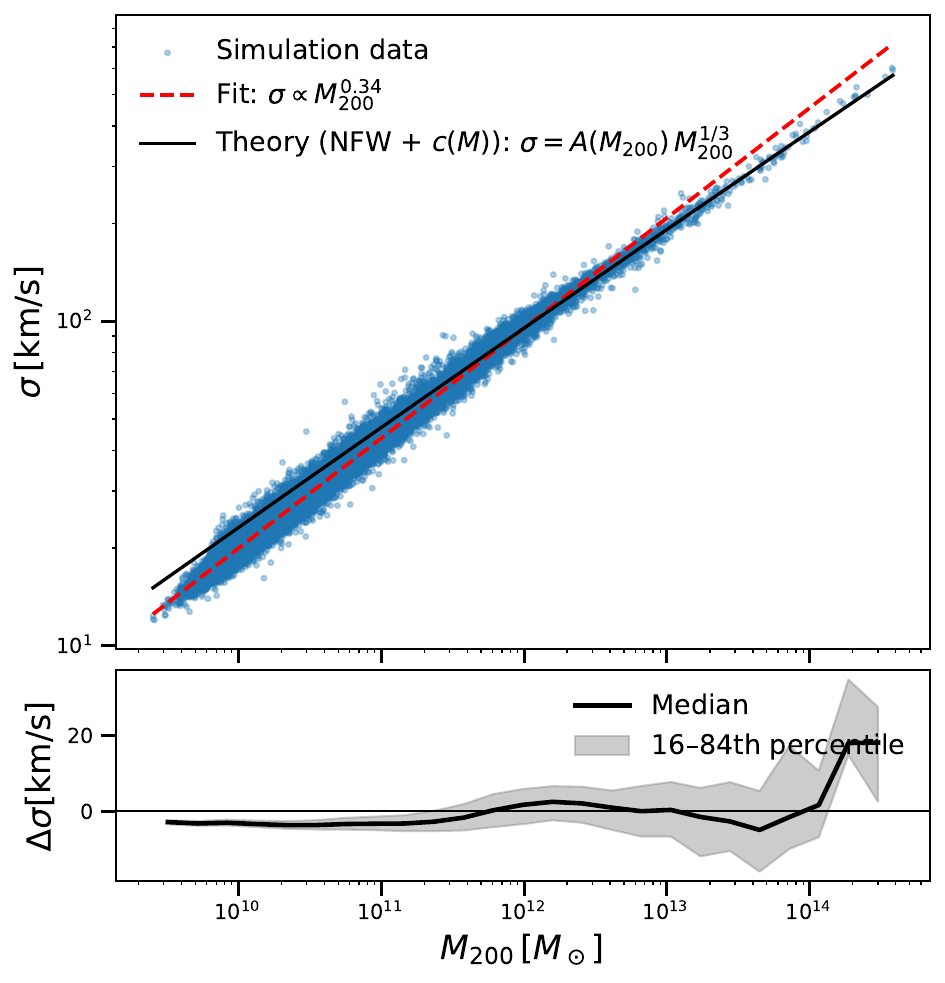}
    \caption{\textit{Relation between velocity dispersion ($\sigma$) and halo mass ($M_{200}$). Each point represents a halo in the sample. The dashed line shows the best-fit power-law relation ($\sigma \propto M_{200}^{0.34}$),while the black solid line shows the theoretical virial expectation for an NFW halo including a mass-dependent concentration $c(M_{200})$from Dutton \& Maccio (2014), such that ($\sigma = A(M_{200})M_{200}^{1/3}$); the curve is normalized to match the simulation at $M_{200}=10^{12}\, M_\odot$. The lower panel shows the residuals ($\Delta\sigma$) relative to the theory, where the black line denotes the median and the gray shaded region indicates the 16th-84th percentile range, corresponding to a typical $1\sigma$ scatter of  $\approx4.52$ km s$^{-1}$, demonstrating that most halos lie close to virial equilibrium. The agreement with the virial expectation is tightest at high halo masses and degrades toward lower masses, where both the scatter and deviations increase. The tighter constraint of the power law fit at low masses reflects the larger number of low-mass halos dominating the regression, while the increasing deviation from the virial expectation toward lower masses indicates deviations from the Dutton \& Macciò (2014) concentration-mass relation. The physical origin of these deviations, likely related to baryonic feedback and resolution effects in TNG100, will be explored in future work. }}
   \label{Virial_relation}
\end{figure}

Figure~\ref{Virial_relation} illustrates the relationship between halo mass and velocity dispersion. From this plot, we see that most galaxies are virialized, though not perfectly so. The small scatter (about 4.52~km~s$^{-1}$) around the line reflects physical differences between systems: some lie above or below, depending on their individual assembly histories, while those right on the line are the most relaxed. In other words, the scatter encodes how close a galaxy is to virial equilibrium. 
We compute the theoretical velocity dispersion assuming a virialized NFW halo. The virialized halo is expected to have a velocity dispersion given by 
\[
\sigma = \left[\frac{G(4\pi\rho_{200})}{3}\right]^{1/3}
\left[\frac{c}{1+c}\left(\ln(1+c) - \frac{c}{1+c}\right)\right]^{1/2}
M_{200}^{1/3},
\]
where the concentration parameter c($M_{200}$) is given by Dutton \& Macciò (2014) as
 \[
    c = 7.85\left(\frac{M_{200}}{3\times10^{12}\,M_\odot}\right)^{-0.081}
    \]
To first order, the NFW term implies $A(M_{200}) \propto c(M_{200})^{1/2}$
. Because $c(M_{200}) \propto M_{200}^{-0.081}$, this yields
$c(M_{200}) \propto M_{200}^{-0.040}$, 
and therefore an effective virial scaling 
$\sigma(M_{200}) \propto M_{200}^{1/3 - 0.0405} \approx M_{200}^{0.29}$.
Thus, even for virialized halos, the combination of an NFW profile and mass-dependent concentration produces a velocity-mass relation that is slightly shallower than the idealized $M_{200}^{1/3}$.  

At the same time, our statistical machine learning problem is complicated by the presence of two variables that are tightly correlated.
\begin{figure}[h]
    \centering
    \includegraphics[width=1\textwidth]{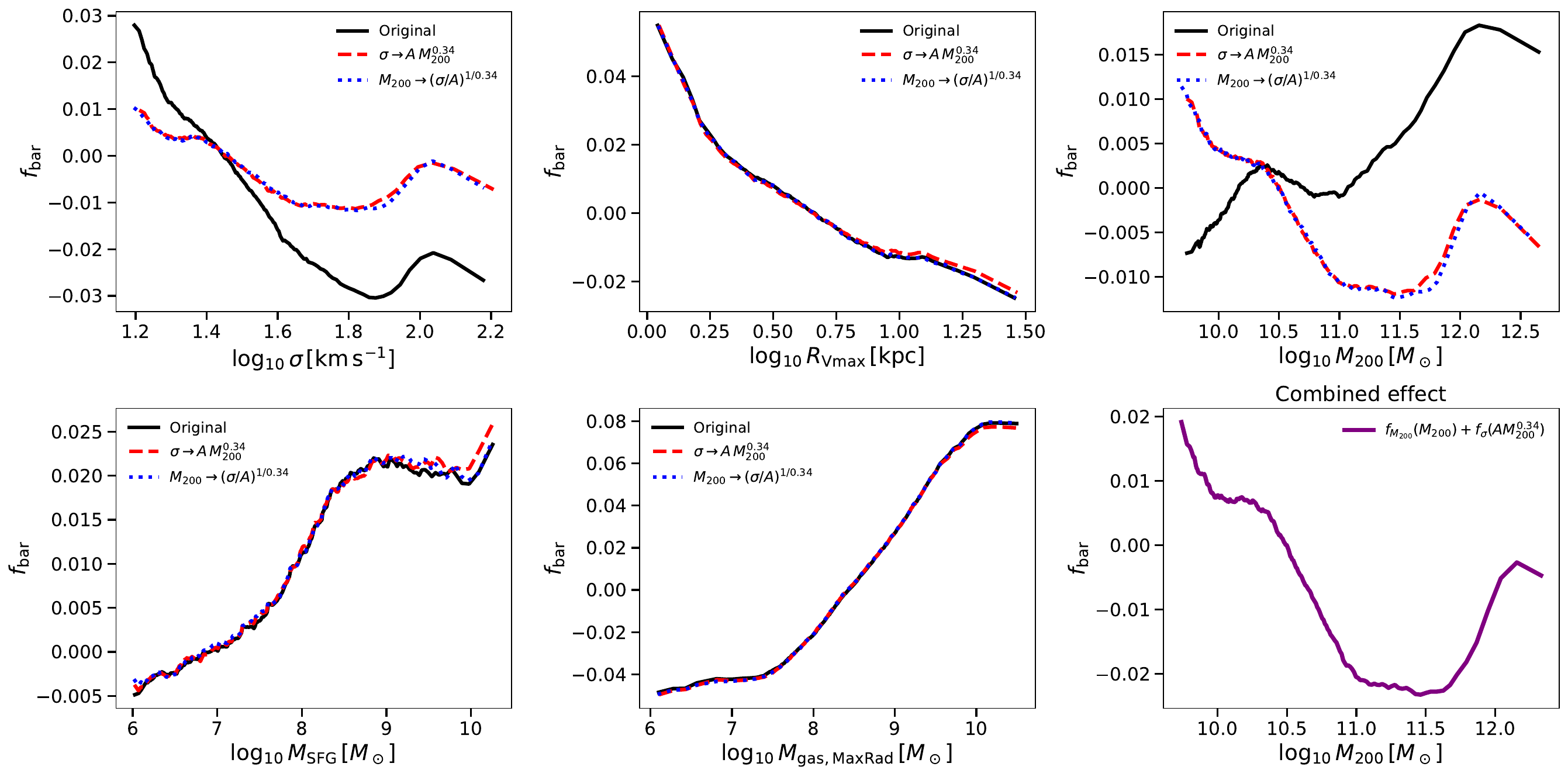}
    \caption{\textit{EBM univariate functions for the five important features after replacing correlated inputs with their virial proxies. Black curves show the original EBM model, the red dashed curves show the substitution $\sigma \rightarrow A M_{200}^{\beta}$, \quad \text{where } $\beta = 0.34$ \text{is the virial slope}, and the blue dotted curves show the substitution $M_{200} \rightarrow \left(\frac{\sigma}{A}\right)^{1/\beta}$. The bottom-right panel shows the combined contribution of $f_{M_{200}}$ and $f_{\sigma}(A M_{200}^{0.34})$. Predictions nearly identical across three features, with deviations appearing  in the $M_{200}$ and $\sigma$ panels, reflecting the strong correlation between $M_{200}$ and $\sigma$.}}
    \label{univariate_overlapping_plots}
 \end{figure}
To explore the effect of highly correlated inputs on our machine-learning model, we replaced one variable with its virial proxy at a time: 
\[
\begin{aligned}
\text{(i)}\quad & \sigma \rightarrow A\,M_{200}^{0.34}  \ \text{(red dashed)} &&\Rightarrow\ R^{2} = 0.864 \pm 0.0009 \\
\text{(ii)}\quad & M_{200} \rightarrow (\sigma/A)^{1/0.34} \ \text{(blue dotted)} &&\Rightarrow\ R^{2} = 0.861 \pm 0.001 \\
& \text{compared to the original model (black):} && R^{2} = 0.866 \pm 0.001
\end{aligned}
\]

Across three of the five feature panels, the predicted baryon fraction hardly changes, but the $M_{200}$ and $\sigma$ panels show changes (Figure~\ref{univariate_overlapping_plots}). The bottom-right panel shows the combined additive contribution of $f_{M_{200}}$ and $f_{\sigma}(A M_{200}^{0.34})$. Despite changes in individual curves, the combined effect remains nearly unchanged. This indicates information leakage between $M_{200}$ and $\sigma$: because they are strongly correlated, replacing one with a proxy doesn’t degrade performance much, but the model’s univariate curves do change. We can interpret this; the two variables encode essentially the same underlying physical information; through the virial relation, they act as proxies for one another, so the model can trade off or mix their contributions without changing its total prediction.
\begin{figure}[h]
    \centering
    \includegraphics[width=0.8\textwidth]{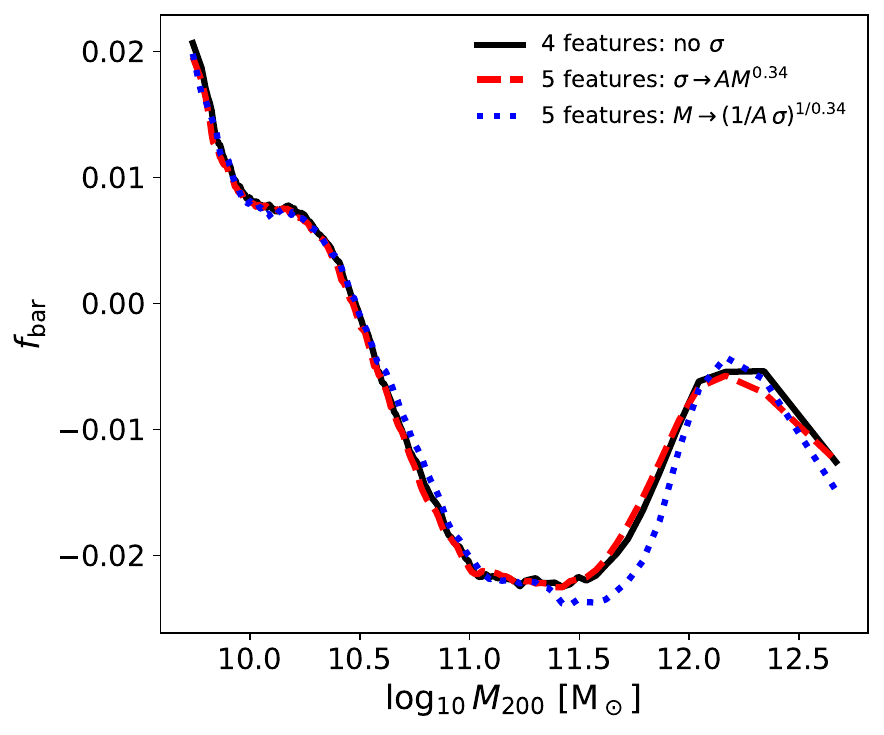}
    \caption{\textit{EBM univariate functions after retraining without velocity dispersion. Black: original EBM trained with four input features ($M_{\text{gas, MaxRad}}$, $M_{SFG}$, $M_{200}$, and $R_{V_{Max}}$); red-dashed: EBM trained with five input features where $\sigma$ is replaced with $\sigma \rightarrow A M_{200}^{\beta}$; blue dotted: EBM trained with five input features, where $M_{200}$ is replaced by the proxy  $M_{200} \rightarrow \left(\frac{\sigma}{A}\right)^{1/\beta}$. Predictions remain unchanged, and the proxy curves show that the model mixes information from these strongly correlated variables.}}
    \label{four_feature_plots}
 \end{figure}

Even after retraining the model using only the four features with $\sigma$ removed, the predictions are effectively the same (Figure~\ref{four_feature_plots}). This is because $\sigma$ maps onto $M_{200}$ through virial relation and does not add independent signal beyond mass. Removing $\sigma$ leads the model to reassign its contribution primarily to $M_{200}$, producing only changes in the univariate response while preserving the overall predictions. The close match between $M_{200}$ and the proxy curves of $M_{200}$ and $\sigma$ show that the model is mixing information from these strongly correlated variables rather than treating one as uniquely causal.
\begin{figure}[h]
    \centering
    \includegraphics[width=1\textwidth]{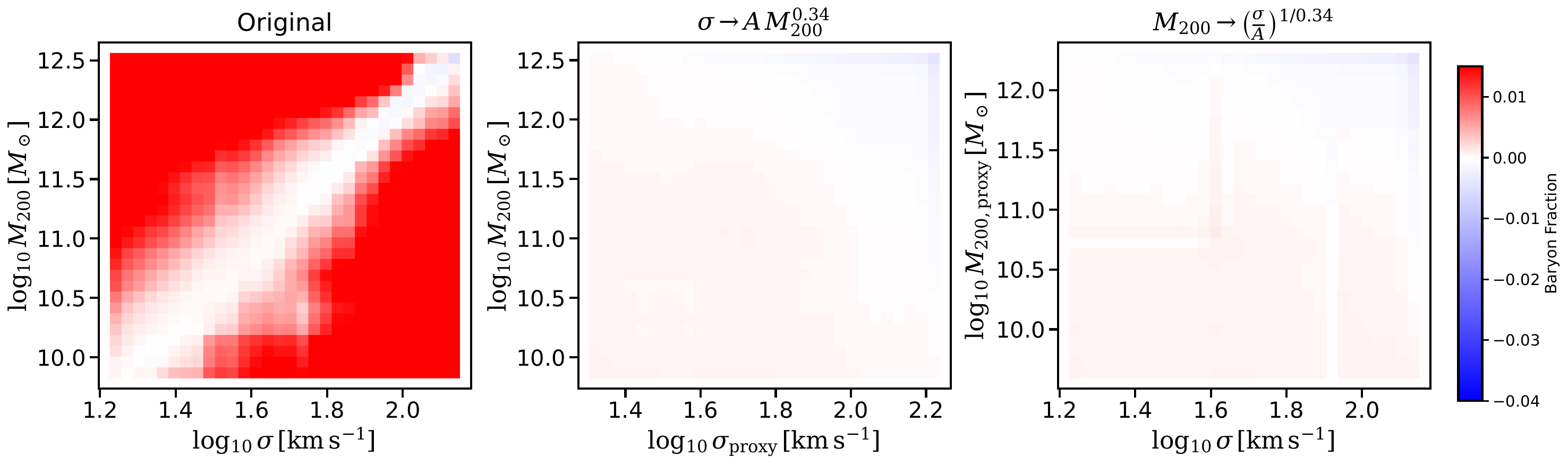}
    \caption{\textit{Bivariate EBM predictions for baryon fraction using $M_{200}$ and $\sigma$(left) and their virial proxies (middle, right). The proxy substitutions remove independent variation, adding no independent signal beyond other variable.}}
    \label{ebm_bivariate_proxies}
 \end{figure}

Figure~\ref{ebm_bivariate_proxies} presents the bivariate heatmaps of the baryon fraction predicted by the model as a function of the halo mass and velocity dispersion (or their proxies) simultaneously. In the original case, a clear diagonal trend appears, indicating a strong correlation between the two variables. After replacing one with its virial proxy, it does not carry additional independent information beyond the other variable.

\section{Discussion}
\subsection{Machine Learning Model Insights and Predictive Performance}
Employing Random Forest and EBM models, our analysis demonstrates that the large-scale galaxy-retained baryon fraction can be reasonably predicted with only five features. The main findings of this work are highlighted in Figures~\ref{scatterplots_RF_EBM}, \ref{univariate_features_importance}, and \ref{bivariate_features_importance}.

By restricting the predictors to the set of the five most important variables identified by the Random Forest model and trained with the EBM model to infer the connections between features and baryon fractions, we were able to train computationally cheap but interpretable models with good predictive performance.

The difference between the training and testing MAE suggests some overfitting during the model-training process, which is expected to some extent. This overfitting was slightly more pronounced in the Random Forest model compared to the EBM model, as indicated by a larger gap between the training and testing errors. However, the impact on overall performance was minimal, as the testing error for the Random Forest model remained within at $<10\%$, and the model's predictive accuracy on unseen data did not degrade significantly. As the Random Forest model was primarily used for feature selection, the minimal overfitting observed was deemed acceptable and thus disregarded. In contrast, no overfitting issues were observed in the case of the EBM model. 

The decrease we see in our raw scatterplots in the $M_{200} \approx 10^{12}\text{--}10^{12.5}\,M_{\odot}$ range  mirrors the behavior reported by \cite{wright2024}, indicating that our analysis captures the same physical regime of baryon loss. The shape learned by the EBM also rises from $\log_{10} M_{200} \sim 11$ to a peak near $\log_{10} M_{200} \sim 12$ and then declines, consistent with \cite{wright2024}. At the highest masses, however, the model does not recover the full scale, which is due to the limited number of massive haloes and the quantile binning that can under-resolve sparse regions of parameter space. In the low-mass regime ($\log_{10}M_{200}\lesssim 11$), the trough in the EBM univariate function represents the isolated marginal contribution of $M_{200}$ after accounting for correlated predictors, rather than the conditional medain baryon fraction at fixed halo mass; the full model prediction remain consistent with the binned raw data in this regime. Consequently, the apparent discrepancy with \cite{wright2024} at low halo masses arises from differences in our halo selection and modeling choices, specifically, the use of quantile binning and the inclusion of lower mass haloes ($M_{200} \geq 4.5 \times 10^9\, M_\odot$) in our data set rather than a new physical effect.

The findings of bivariate feature functions showed how the gas mass, halo mass, and velocity dispersion influence baryon retention in galaxies. 
If galaxies are in virial equilibrium, then $M_{200} \propto \sigma^{3}$ is the expected relationship between $M_{200}$ and $\sigma$. However, the learned interaction between $M_{200}$ and $\sigma$ shows deviations from this theoretical scaling, indicating that galaxies outside the virial equilibrium tend to retain more baryons. Additionally, the interaction between $M_{\text{gas, MaxRad}}$ and $M_{200}$ suggests that gas-poor galaxies have low baryon fractions due to weak gravitational potential. High-mass halos retain baryons even with low central gas, while those with both high halo mass and central gas retain more baryons.

\subsection{Comparison to Previous Studies}

\cite{jo2019} used a machine-learning pipeline based on the ExtraTreeRegressor (\cite{Geurts2006}) to predict galaxies’ baryonic properties (e.g., gas mass, stellar mass, black hole mass, SFR, metallicity and stellar magnitude) based on dark matter (DM) halo features, such as DM mass, velocity dispersion, maximum circular velocity, angular momentum, merger history, and number of nearby halos. Their model, trained on IllustrisTNG100-1 simulation data, identified key correlations between DM and baryonic features, with varying feature importance depending on redshift. They found maximum circular velocity was critical at high redshift, while halo mass and velocity dispersion were more important at lower redshift.
In contrast, our analysis identified halo mass and velocity dispersion along with $M_{\text{gas, MaxRad}}$, $M_{SFG}$, and $R_{V_{Max}}$ as the most important predictors for baryon retention.
 
\cite{Ayromlou2023} utilized three sets of cosmological hydrodynamical simulations—IllustrisTNG50 (\cite{nelson2019};\cite{pillepich2019}, TNG100 (\cite{Springel2018};\cite{naiman2018}), TNG300 (\cite{marinacci2018};\cite{Nelson2018}), EAGLE (\cite{Crain2015};\cite{schaye2015}), and SIMBA (\cite{dave2019})— to study baryon distribution in and around halos. They showed that baryonic feedback significantly impacts gas redistribution. TNG and EAGLE models exhibit similar trends in halo baryon fractions as a function of halo mass, aligning well with X-ray and SZ observations for the most and least massive halos. In contrast, SIMBA predicted stronger AGN feedback, ejecting more baryons to larger distances and resulting in higher baryon fractions far from the halo center compared to TNG and EAGLE. TNG and EAGLE predict nearly identical closure radii, the distance from a halo's center where the integrated baryon fraction reaches the cosmic average. 

In our analysis, we found halo mass to be an important predictor of baryon fractions. Unlike \cite{Ayromlou2023} who analyzed baryon fractions at different distances from the halo center, we focused on single, integrated values for the entire halo. Our results highlight correlations between baryon fractions and parameters, such as $M_{\text{gas, MaxRad}}$, $M_{\text{SFG}}$, $M_{200}$, $R_{V_{Max}}$, and $\sigma$. We also observed a decrease in baryon fractions with increasing $R_{V_\text{Max}}$, a factor not addressed in their work, which instead relied on closure radii to study baryon distributions.

Our approach also shares similarities to \cite{ Hausen2023} who used an EBM model to analyze data from CROC simulations. They used galactic environment and DM halo properties to predict the $M_{*}$ and $SFR$. While their study targeted star formation rate and stellar mass, our focus is on the total baryonic content of galaxies. They used EBM to reveal the relative importance of galactic properties in setting $M_{*}$ and $SFR$, while we employ a multi-step interpretable machine learning framework. Our approach not only ranks the important features, but also infers the relationship between baryon fraction and galaxy properties, providing a more comprehensive analysis.

\cite{romeo2020} showed that a scaling relation involving stellar specific angular momentum ($j*$) and stellar radial velocity dispersion ($\sigma*$), both linked to disc gravitational instability, provides a much tighter constraint on the galaxy-halo connection than traditional stellar-to-halo mass relations. In our simulation-machine learning-based analysis, virial scaling variables such as $M_{200}$ and $\sigma$ along with $M_{\text{gas, MaxRad}}$, $M_{SFG}$, and $R_{V_{Max}}$ emerge as significant predictors of baryon retention, suggesting a possible link to the dynamical equilibrium framework proposed by \cite{romeo2020}.

\subsection{Limitations/Future Works}
The primary limitation, and the possible extension, of our work is the use of a single simulation. Applying this approach to multiple simulations, such as EAGLE, SIMBA etc., can help determine how well it generalizes. This could aid in determining why these models disagree as to the states of their CGM, as identified by \cite{wright2024}. Our study has focused on the z=0 properties of galaxies in TNG100. To gain a deeper understanding of galaxy evolution, we plan to revisit this study at higher redshifts. This will allow us to examine whether the trends in baryon retention we found here hold during earlier epochs of galaxy formation. 

In our research, we did not consider variations in feedback parameters, but it would be interesting to explore how these parameters affect our results. One approach could be to use the CAMELS suite of simulations (\cite{villaescusa2021}) et al., which includes over 2000 hydrodynamical simulations with different feedback and cosmology parameters. Alternatively, we could use semi-analytic models (SAMs), which are computationally efficient for modeling galaxy formation (\cite{Somerville2008}).

Our machine-learning models capture correlations between halo or galaxy properties and baryon fractions but do not establish causal relationships. For instance, while we observe that higher $M_{200}$ correlates with higher retained baryon fractions, we cannot conclude that increasing halo mass directly causes this outcome. To understand the direction of cause and effect, future work could analyze the temporal evolution of individual galaxies using multiple simulation snapshots, which would allow us to track how changes in feedback, mass, and gas content develop over time. This time-resolved approach could help clarify whether halo growth, feedback processes, or other factors drive certain trends.

 \section{Conclusions}
In this study, we used interpretable multi-step machine learning to discover what galaxy properties determine the retention of baryons. 

We concentrated on various galactic, halo, and dynamical properties for predicting the retained baryons.  In the first step, we identified the most significant features for predicting the retained baryon fraction using Random Forest Regressor. Then, we used the EBM model to understand how these features contribute to predicting the retained baryon fractions. 

Here are the major findings summarized:
\begin{enumerate}
    \item Both techniques proved to be highly accurate, with Random Forest achieving nearly 89.7\% $R^2$ score and EBM approximately 86.6\%. Additionally, reducing the features in the EBM from 66 to 5 had a minimal effect on its accuracy.

    \item The Random Forest model effectively identifies the top five features, highlighting $M_{\text{gas, MaxRad}}$, $M_{\text{SFG}}$ and $M_{200}$ as key halo properties for predicting baryon fractions. $M_{\text{gas, MaxRad}}$ (gas mass within the radius of $V_{Max}$) is particularly crucial, with an importance score of 1.799.
    
    \item The univariate functions $M_{\text{gas, MaxRad}}$ and $M_{\text{SFG}}$  are positively correlated with the baryon fraction, whereas ($R_{V_{Max}}$) is negatively correlated. This indicates that compact galaxies with a higher abundance of star-forming gas and a greater gas mass within the radius of $V_{Max}$ are associated with higher baryon fractions.
    
    \item Our analysis highlights that $M_{200}$ reveals a non-monotonic relationship between $M_{200}$ and baryon fraction: larger haloes tend to keep more baryon overall but show a decline at high masses, likely linked to AGN-driven outflows removing gas. We also find that $M_{200}$ and $\sigma$ are strongly correlated through the virial relation that the EBM can exchange contributions without changing its predictions. This tells that the two features encode the same underlying gravitational potential and act as proxies for one another rather than representing two independent drivers. 
    
   \item Baryon retention increases with the interaction between $M_{200}$ and $\sigma$, showing that galaxies outside the virial equilibrium retain more baryons. The interaction between $M_{\text{gas, MaxRad}}$ and $M_{200}$ shows that high-mass halos can retain baryons despite low gas content, while low-mass gas-poor halos lose baryons. 
\end{enumerate}

\section*{Acknowledgments}

We acknowledge the support from the Department of Physics and Materials Science at the University of Memphis, Tennessee. We are incredibly grateful to KM Ashraf, whose generous sharing of his machine-learning expertise has been instrumental in our research. I also want to express my gratitude for the
unwavering support of my loving family. Additionally, we thank the IllustrisTNG team for their efforts in developing the IllustrisTNG simulation code. Support for program HST-AR-17547  was provided by NASA through a grant from the Space Telescope Science Institute, which is operated by the Associations of Universities for Research in Astronomy, Incorporated, under NASA contract NAS5-26555.

We acknowledge the use of the following open-source software packages in our analysis: numpy (\cite{harris2020}), pandas (\cite{mckinney2010}), matplotlib (\cite{hunter2007}), scikit-learn (\cite{pedregosa2011}), seaborn (\cite{waskom2021}), interpret (\cite{nori2019}), and h5py (\cite{collette2013})

\section{Data Availability}
The data directly supporting the findings of this work are available from the corresponding author upon request. The simulation data used in this work are from the publicly available IllustrisTNG project and can be accessed at \url{https://www.tng-project.org/} (\cite{nelson2019}).
All analysis scripts and machine learning code used in this study are available on GitHub at: \url{https://github.com/mkhanom/Baryon\_Fraction\_ML}

\bibliographystyle{aasjournal}
\bibliography{references}{}

\appendix
\section{Robustness of the EBM model} 

We performed a thorough evaluation by using many random subsets of our data ensuring the robustness of our EBM model. We divided the data into ten random subsets, each consisting of training and testing sets. We provided training on the EBM model for each splitting and assessed its performance by calculating the $R^2$ score. Furthermore, we calculated the importance of the top 5 features in each model.

\begin{figure}
    \centering
    \includegraphics[width=1\textwidth]{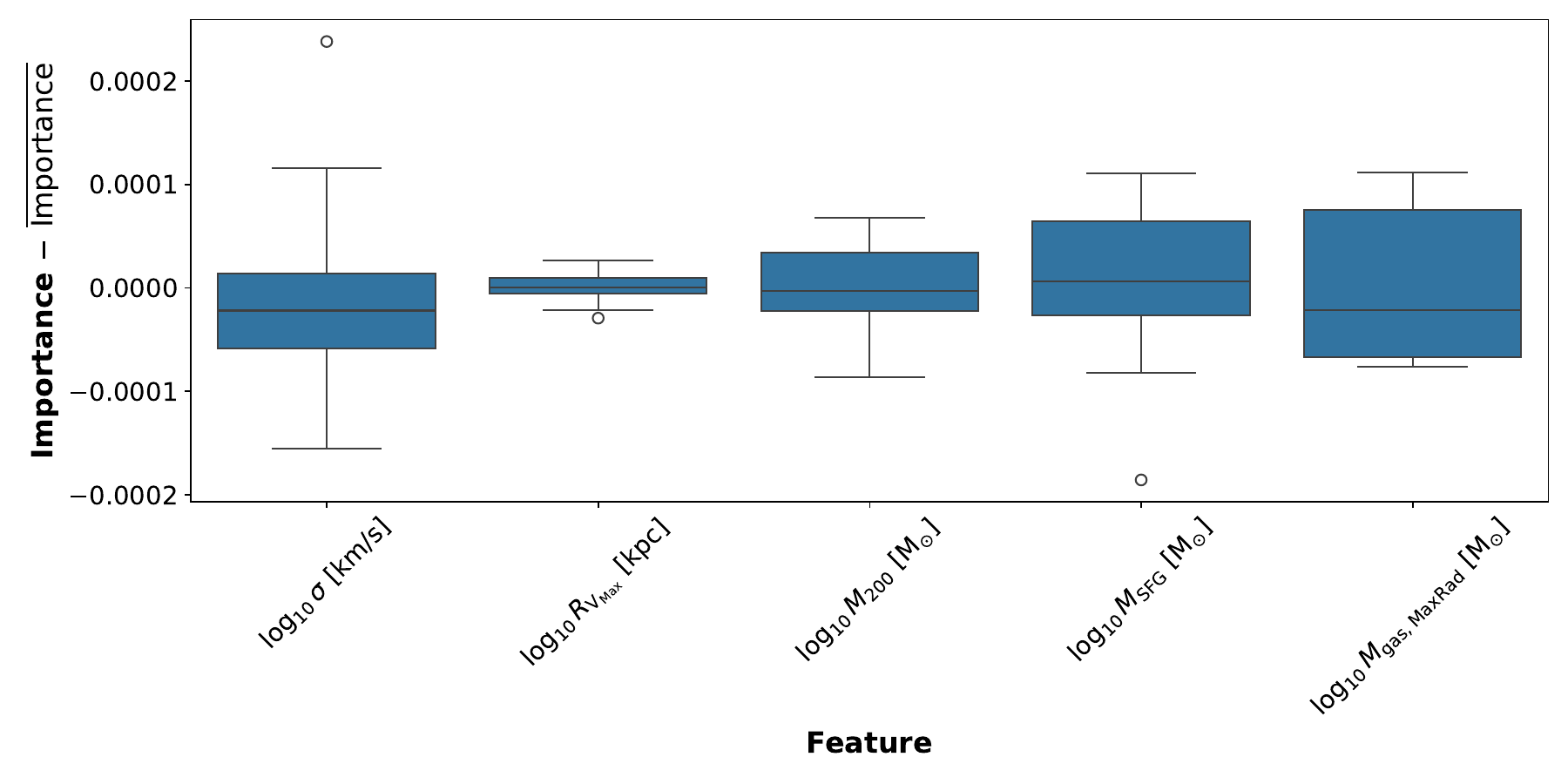} 
    \caption{\textit{The variation in relative importance (\(\text{Importance} - \overline{\text{Importance}}\)) of the top 5 features for 10 different random subsets based on standard deviation values.}}
    \label{fig:univariate_feature_std}
 \end{figure}

Figure~\ref{fig:univariate_feature_std} shows the variability of each feature's importance values of the top 5 features across these 10 distinct random subsets using boxplots. From this plot, we can see that features such as $R_{V_{Max}}$ and $M_{200}$ exhibit shorter whiskers and a narrower interquartile range, indicating consistent contributions to the model across different subsamples. In contrast, features like $M_{\text{gas, MaxRad}}$ and $M_{\text{SFG}}$  show slightly higher variability, as evidenced by their wider boxes and occasional outliers. However, the overall low spread in the centered values across all the features in different subsets underscores the reliability of these feature importances.

\begin{figure}
    \centering
    \includegraphics[width=1\textwidth]{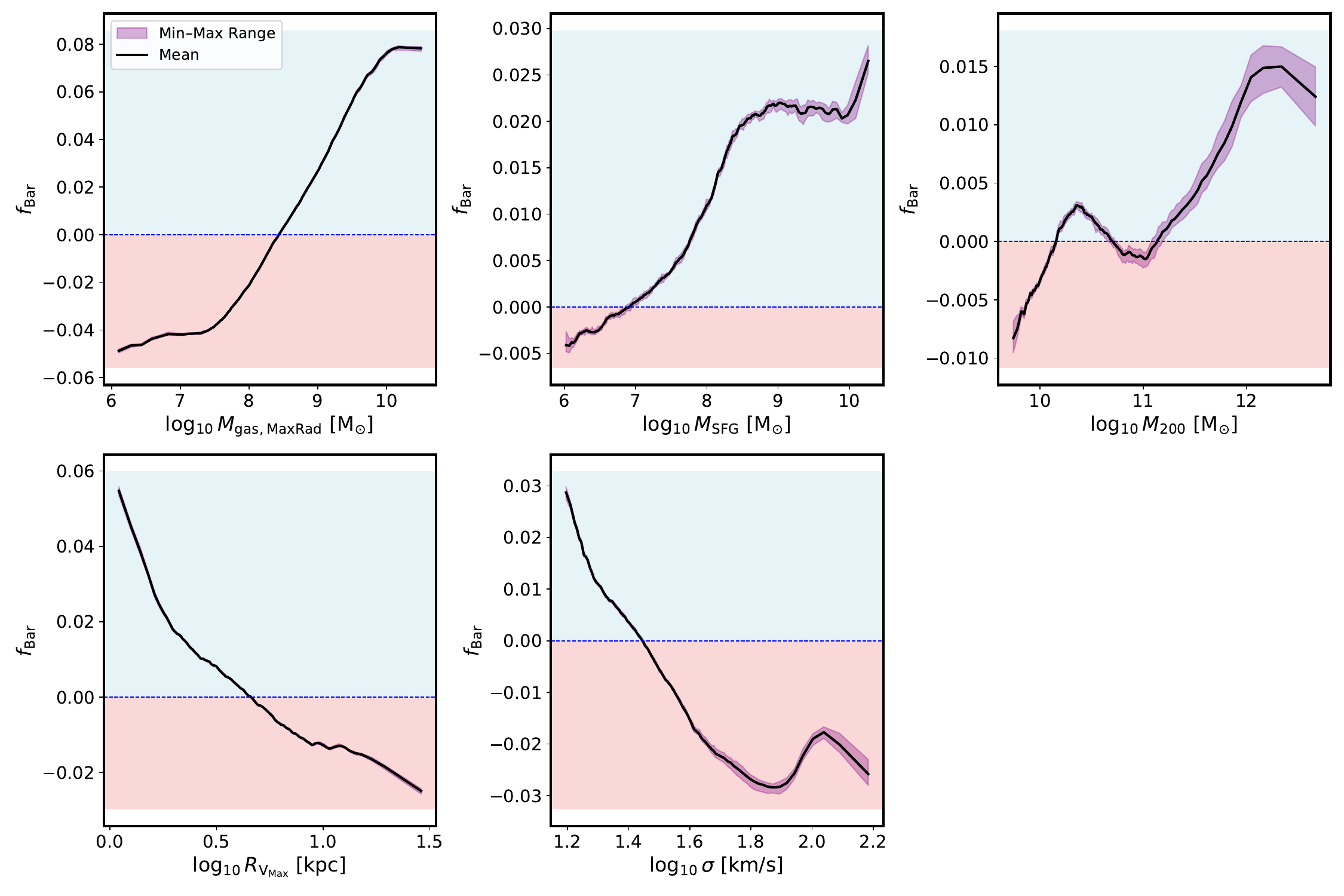}
    \caption{\textit{Univariate feature functions for predicting the baryon fraction across 10 randomly selected subsets. The black line in each panel represents the mean feature effect across 10 random subsets, while the shaded region captures the full range (min to max) among the models. Both the target variable (baryon fraction) and the input features are represented in a logarithmic scale with a base of 10. The overlapping colors represent the 10 different models for each feature.}}
    \label{univariate_feature_importance_multiple_subsets}
 \end{figure}

 We also performed other analyses, such as assessing the stability of feature importance across different random subsets based on standard deviation, the feature importance of univariate feature functions for different random subsets, the standard deviation for bivariate feature functions, and computing z-scores for each feature:

 Figure~\ref{univariate_feature_importance_multiple_subsets} illustrates the individual feature functions of the EBM model focused on predicting baryon fraction across 10 random subsets. Notably, the plot reveals minimal fluctuation in feature importance among these subsets. The consistency in feature importance across diverse data subsets suggests that the model is robust for all features except $M_{200}$ and $\sigma$, which exhibit some deviations across the different subsets. 

\begin{figure}
    \centering
    \includegraphics[width=1\textwidth]{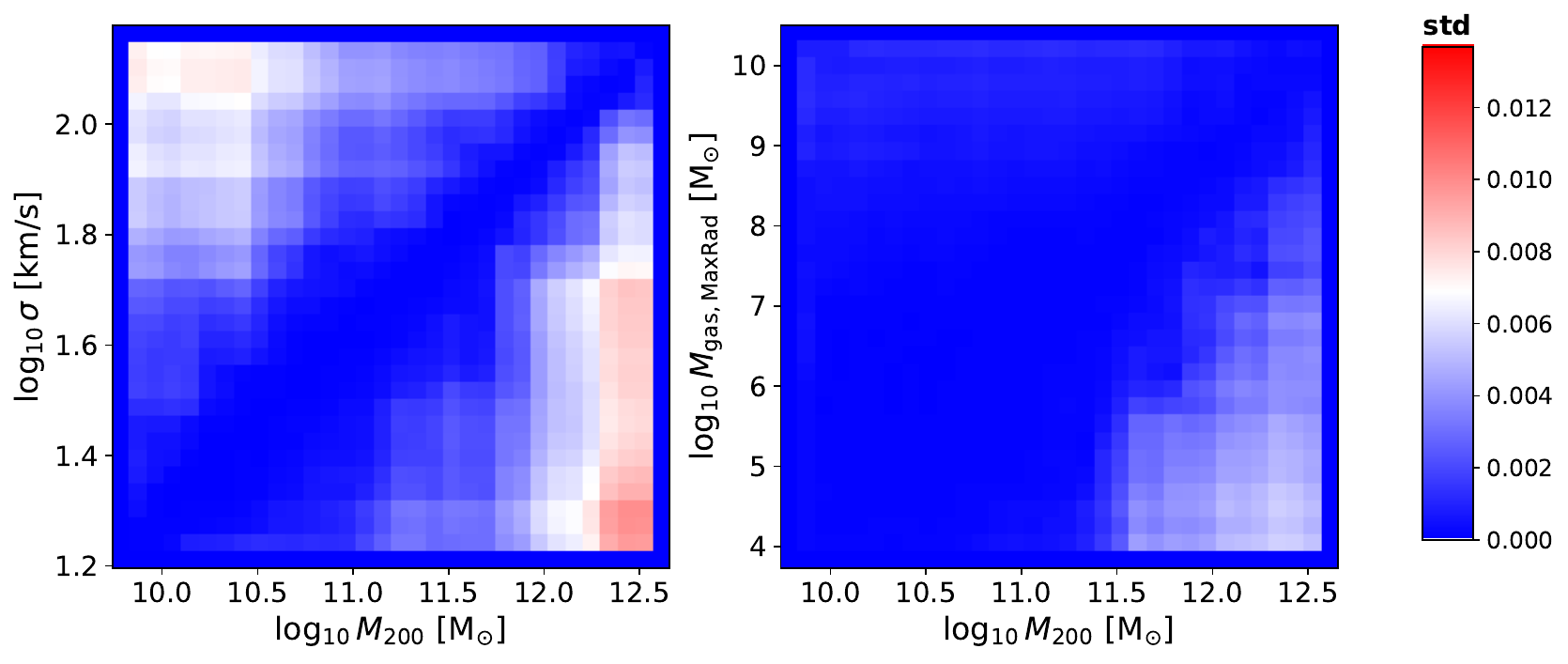}
    \caption{\textit{The standard deviation maps of bivariate interaction terms from the EBM were calculated across ten random train/test splits. Each panel represents the variability in interaction effect between a pair of features, as learned by the model in predicting baryon fractions in galaxies. Red regions with high standard deviation indicate greater variability in the interaction effect across models, pointing to areas of model sensitivity or uncertainty. Dark blue areas reflect regions, where feature interactions have no variation.}}
    \label{bivariate_feature_std}
 \end{figure}

Figure~\ref{bivariate_feature_std} depicts the standard deviation of model scores for bivariate features.

 From this plot, we can see that most of the feature pairs exhibit no variation in the model's predictions. However, the combinations involving high $M_{200}$ with low to medium $\sigma$, and low $M_{\text{gas, MaxRad}}$ with high $M_{200}$ show minimal variability, indicating that these feature interactions contribute consistently to the model's output.
 
 \begin{figure}
    \centering
    \includegraphics[width=1\textwidth]{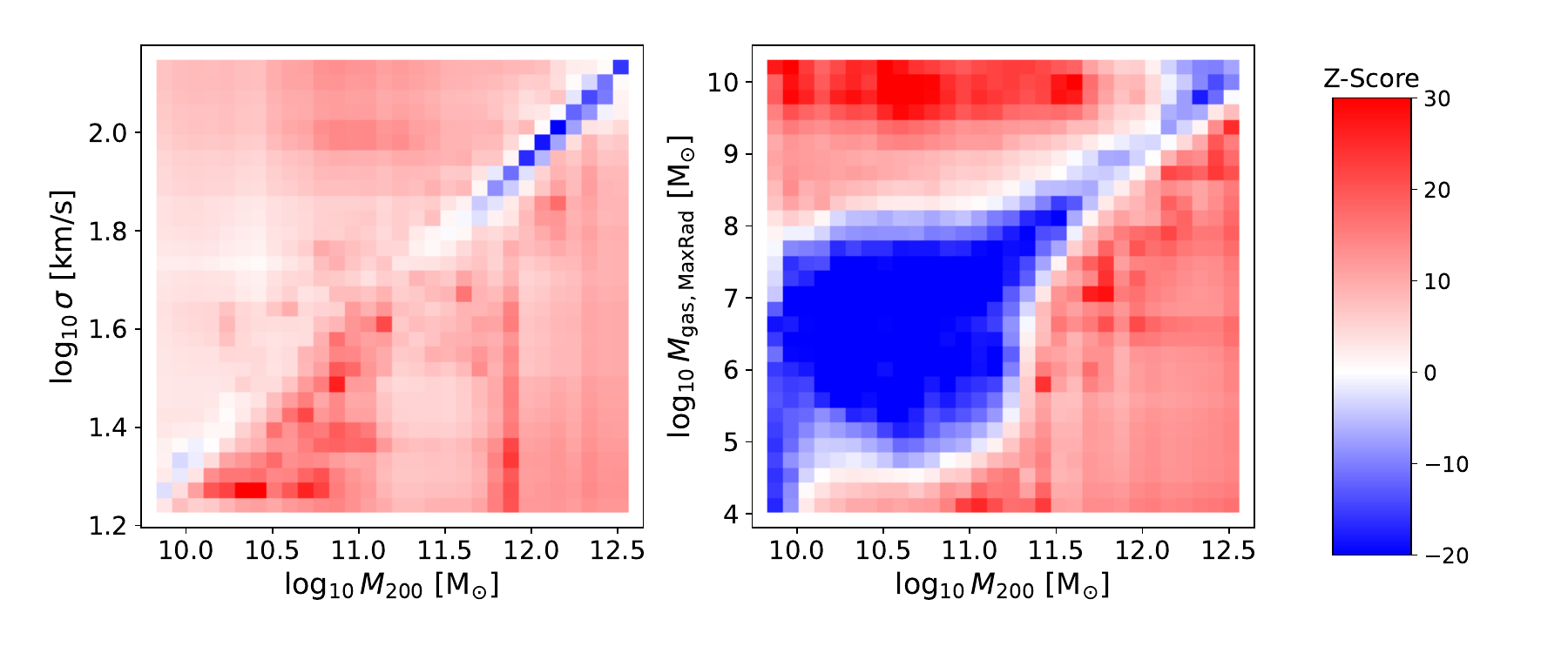}
    \caption{\textit{Z-scores of bivariate interaction terms from the EBM model, computed across ten random train/test splits. Each panel shows a pair of features, with z-scores indicating how consistently and strongly the interaction contributes to the model’s prediction of baryon fraction. Red regions indicate statistically robust positive contributions; blue regions indicate robust negative contributions, and white areas correspond to weak or inconsistent effects.}}
    \label{z_score}
 \end{figure}

Figure~\ref{z_score} shows the z-scores for bivariate functions from the EBM model, which were calculated over ten different random train/test splits. The z-scores indicate how consistently and strongly that specific region of the bivariate interaction contributes (positively or negatively) to the model’s prediction across multiple models. A high positive z-score (red) in that region consistently contributes positively to predicting higher baryon fractions across models. A high negative z-score (blue) indicates that region consistently contributes negatively to the prediction across models. White regions indicate interactions that are weak or statistically inconsistent across models.

For example, the interaction between $\log_{10}M_{200}$ and $\log_{10} \sigma$ exhibits positive z-scores across most regions, highlighting a consistently positive contribution to baryon retention predictions across, except when both halo mass and velocity dispersion are high, where the contribution becomes negative (blue). Similarly, the interaction between $\log_{10}M_{200}$ and $\log_{10} M_{\text{gas, MaxRad}}$ reveals robust regions of both positive and negative influence, indicating that the model detects statistically stable patterns in how gas content and halo mass affect the baryon fraction.

\section{Overview of training features and correlations}
 
As previously mentioned, our data set comprises 89 features representing 107, 867 simulated galaxies. Table 7 provides a detailed overview of 84 features, including their symbols, descriptions, and units, while the 5  most  significant features are highlighted in section 2.8, as shown in Table 4.
 
\begin{longtable}{lll}
\caption{Summary of galactic input parameters used for predicting the retained baryon fraction with Random Forest and EBM models} \\
\toprule
Symbol & Description & Unit \\
\midrule
\endfirsthead
\toprule
Symbol & Description & Units \\
\midrule
\endhead
\midrule
\endfoot
\bottomrule
\endlastfoot
$E_{Quasar}$ & Cumulative energy from quasars & $\mathrm{erg}$ \\
$E_{Wind}$ & Cumulative energy from galactic winds & $\mathrm{erg}$ \\
$R_{200}$ & Virial radius & $\mathrm{kpc}$ \\
$B_{\text{disk}}$ & Magnetic field strength within the disk of a galaxy & $\mu G$ \\
$\text{SFR}$ & The sum of each gas cell's individual star formation rate & $M_{\odot}/\text{yr}$ \\
$T_{\text{SF}}$ & The exact time when the star was formed & $\text{Gyr}$ \\
$Z_{*},{Group}$ & Mass-weighted average metallicity of star particles in the group & - \\
$Z_{gas},{Group}$ & Mass-weighted average metallicity of gas particles in the group & - \\
${SFR}_{Group}$ & Star formation rate of all gas cells within the group & $\mathrm{M_{\odot}\,yr^{-1}}$ \\
$f_{\mathrm{[X]gas},\mathrm{Group}}$ & Fraction of gas cells of species $X$ within the group & - \\
$f_{\mathrm{[X]gas}}$ & Fraction of gas cells of species $X$ within the subhalo & - \\
$f_{\mathrm{[X]*,Group}}$ & Fraction of stars of species $X$ within the group & - \\
$f_{\mathrm{[X]*}}$ & Fraction of stars of species $X$ within the subhalo & - \\
$M_{Group}$ & Total mass of the group & $\mathrm{M_{\odot}}$ \\
$\dot{M}_{\text{BH,Group}}$ & Accretion rate onto black holes within the group & $\mathrm{M_{\odot}\,yr^{-1}}$ \\
$M_{BH,Group}$ & Total mass of black holes within the group & $\mathrm{M_{\odot}}$ \\
$R_{1/2}$ & Radius containing half of the total mass of the subhalo & $\mathrm{kpc}$ \\
$V_{Max}$ & Maximum circular velocity of all particles in the subhalo & $\mathrm{km\,s^{-1}}$ \\
$M_{Rad}$ & Total mass within twice the stellar half-mass radius & $\mathrm{M_{\odot}}$ \\
$M_{R_{1/2}}$ & Total mass within the stellar half-mass radius & $\mathrm{M_{\odot}}$ \\
$M_{MaxRad}$ & Total mass within the radius of maximum velocity & $\mathrm{M_{\odot}}$ \\
$M_{gas,Group}$ & Gas mass of the group & $\mathrm{M_{\odot}}$ \\
$M^{*}_{\text{Group}}$ & Star mass of the group & $\mathrm{M_{\odot}}$ \\
$M_{DM,Group}$ & Dark matter mass of the group & $\mathrm{M_{\odot}}$ \\
$Z_{gas}$ & Gas metallicity within the subhalo & - \\
$Z_{*}$ & Metallicity of the stars within the subhalo & - \\
$Z_{MaxRad}$ & Gas metallicity within the radius of the maximum velocity for the subhalo & - \\
$Z_{*MaxRad}$ & Stellar metallicity within the radius of the maximum velocity for the subhalo & - \\
$Z_{*R_{1/2}}$ & Stellar metallicity within the half-mass radius of the subhalo & - \\
$Z_{gas,R_{1/2}}$ & Gas metallicity within the half mass-radius of the subhalo & - \\
$B_{halo}$ & Magnetic field strength within the halo of the subhalo & $\mu\mathrm{G}$ \\
$Z_{gas,SFR}$ & Metallicity of the gas with star formation within the subhalo & - \\
$R_{\text{gas},\,1/2}$ & Gas half-mass radius & $\mathrm{kpc}$ \\
$R^{*}_{1/2}$ & Star half-mass radius & $\mathrm{kpc}$ \\
$M$ & Total mass of the subhalo & $\mathrm{M_{\odot}}$ \\
$M_{gas}$ & Gas mass of the subhalo & $\mathrm{M_{\odot}}$ \\
$M^{*}$ & Star mass of the subhalo & $\mathrm{M_{\odot}}$ \\
$M_{{DM}}$ & Dark matter mass of the subhalo & $\mathrm{M_{\odot}}$ \\
$M_{{BH}}$ & Mass of the black holes within the subhalo & $\mathrm{M_{\odot}}$ \\
$M^{*}_{\text{Rad}}$ & Star mass within twice the stellar half-mass radius of the subhalo & $\mathrm{M_{\odot}}$ \\
$M_{Rad}$ & Gas mass within twice the stellar half-mass radius of the subhalo & $\mathrm{M_{\odot}}$ \\
$M_{DM, Rad}$ & Dark matter mass Star mass within twice the stellar half-mass radius of the subhalo & $\mathrm{M_{\odot}}$ \\
$M^{*}_{R_{1/2}}$ & Gas mass within the half-mass radius of the subhalo & $\mathrm{M_{\odot}}$ \\
$M_{{1/2}}$ & Star mass within half-mass radius of the subhalo & $\mathrm{M_{\odot}}$ \\
$M_{DM, {1/2}}$ & Dark matter mass within half-mass radius of the subhalo & $\mathrm{M_{\odot}}$ \\
$M^{*}_{\text{MaxRad}}$ & Star mass  within the radius of maximum velocity & $\mathrm{M_{\odot}}$ \\
$M_{DM, {MaxRad}}$ & Dark matter mass  within the radius of maximum velocity & $\mathrm{M_{\odot}}$ \\
$N_{mergers}, {Total}$ & Total number of major mergers & - \\
$N_{mergers}, {LastGyr}$ & Number of major mergers in the last Gyr & - \\
$M^{*}_{\text{ExSitu}}$ & Ex-situ stellar mass & $\mathrm{M_{\odot}}$ \\
$V_{esc}$ & Escape velocity for halo mass and virial radius & $\mathrm{km\,s^{-1}}$ \\
$\phi_{\text{halo mass}}$ & Gravitational potential for halo mass & $(\mathrm{km\,s^{-1}})^{2}$ \\
\end{longtable}

\par \medskip
\textbf{Note:} \( X \) represents the chemical species \(\text{H, He, C, N, O, Ne, Mg, Si, and Fe}\).

Using the Pearson correlation method, we identified feature pairs with strong correlations (correlation coeffcients $> 0.75$). Only strongly correlated feature pairs are presented in the correlation matrix. The correlation matrix for the 27 features is divided into three groups for better visualization. We applied the Variance Inflation Factor (\cite{salmeron2020}) (VIF$<10$) to select features with acceptable multicolinearity. This process led to the removal of 23 features due to significant multicolinearity and leakage issues, retaining 4 features for training the model:
\begin{itemize}
    \item $E_{\mathrm{wind}}$ — Cumulative energy from galactic winds
    \item $M_{\mathrm{gas}, \mathrm{MaxRad}}$ — The gas mass within $R_{V_{\text{Max}}}$
    \item $R_{V_{\mathrm{max}}}$ —The radius of the rotation curve peak ($V_{\text{Max}}$)
    \item SFR — Star formation rate
\end{itemize}

\begin{figure}
    \centering
    \includegraphics[width=1\textwidth]{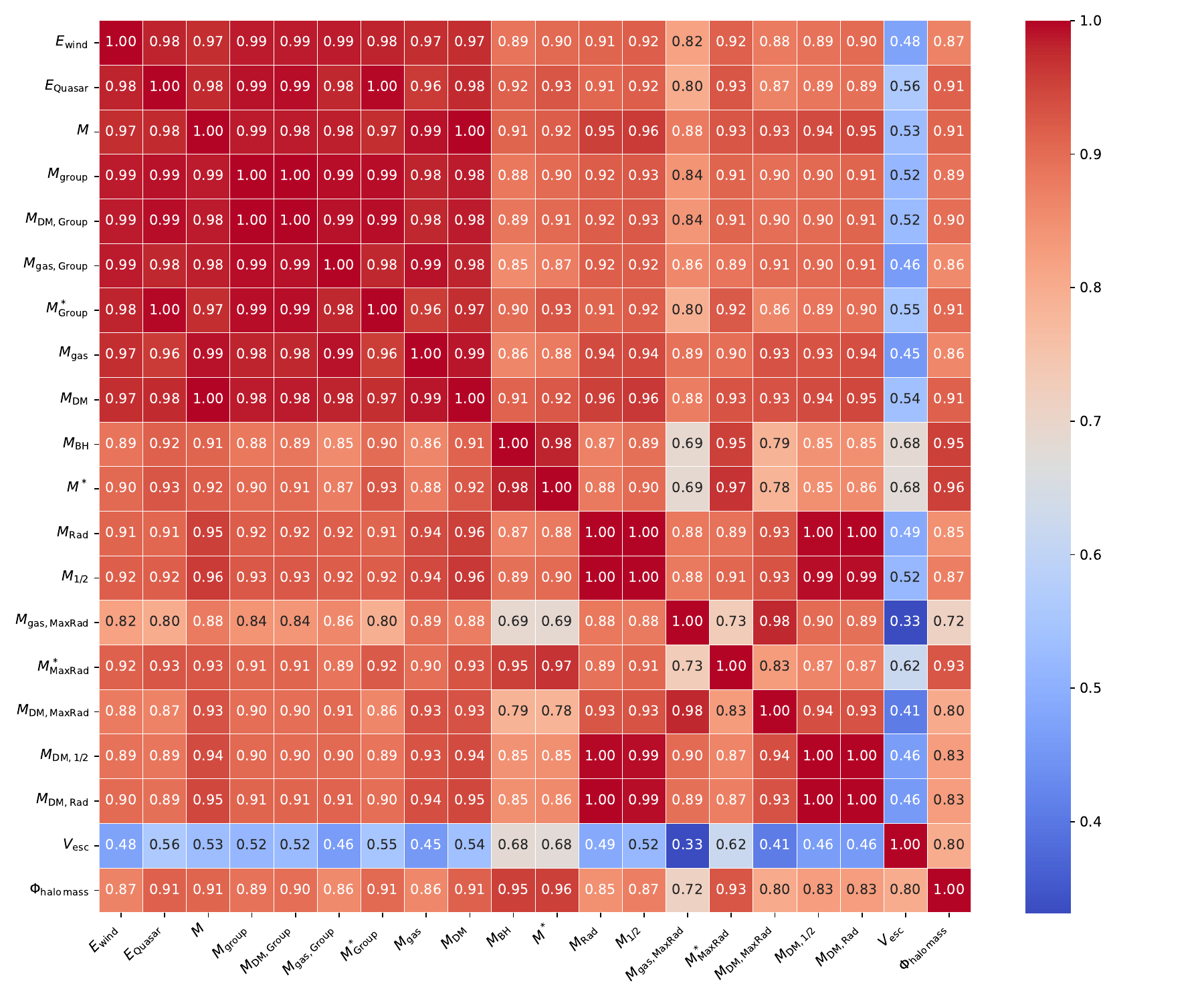}
    \caption{\textit{Correlation matrix of mass and energy features, illustrating the relationships among mass-related properties as well as quasar and wind energy.}}
    \label{corr_mass_energy}
 \end{figure}

 \begin{figure}
    \centering
    \includegraphics[width=1\textwidth]{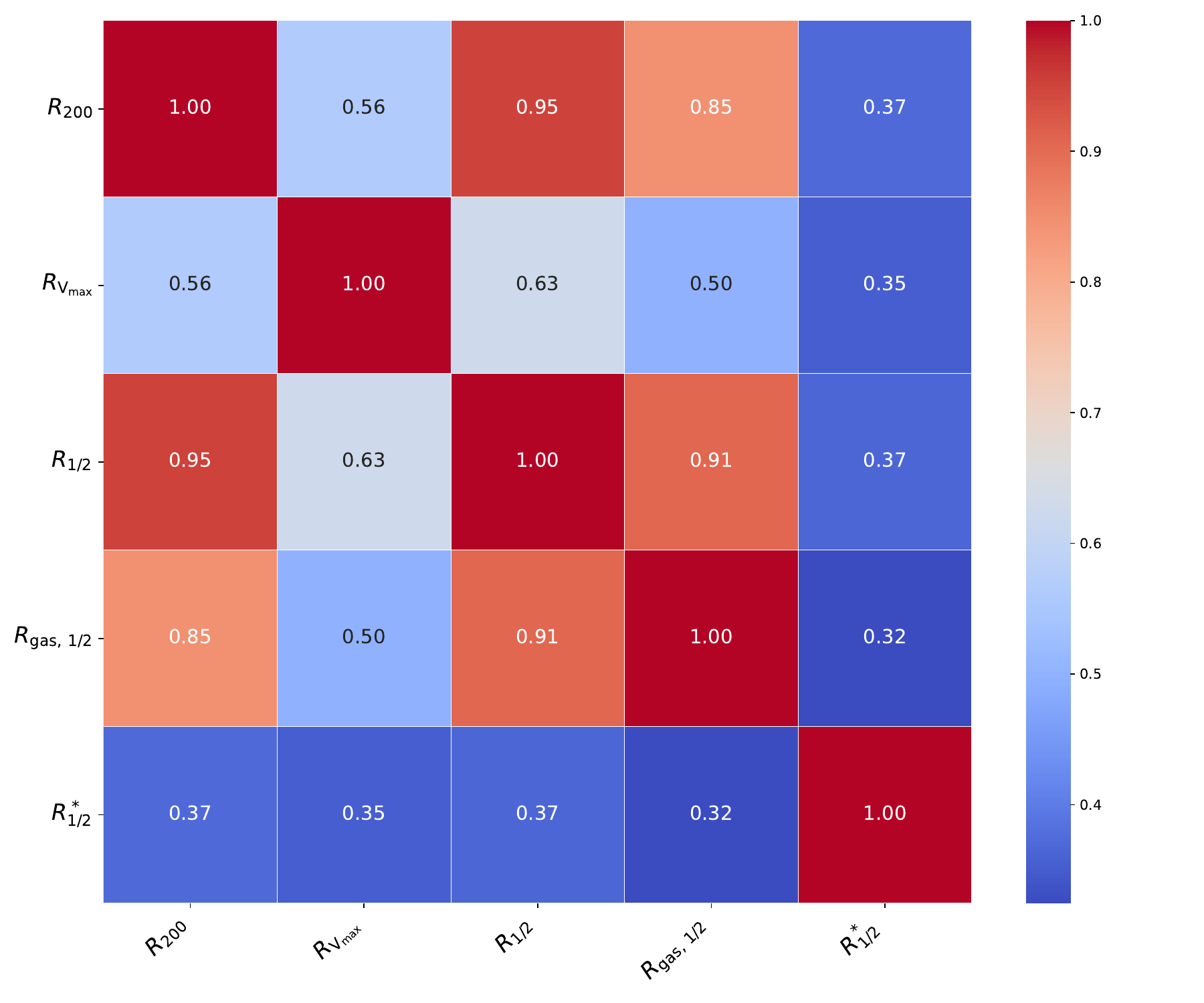}
    \caption{\textit{Correlation matrix for radius features, displaying the relationships between various radii, such as virial radius, half-mass radius, and maximum circular velocity radius.}}
    \label{corr_radius}
 \end{figure}

 \begin{figure}
    \centering
    \includegraphics[width=1\textwidth]{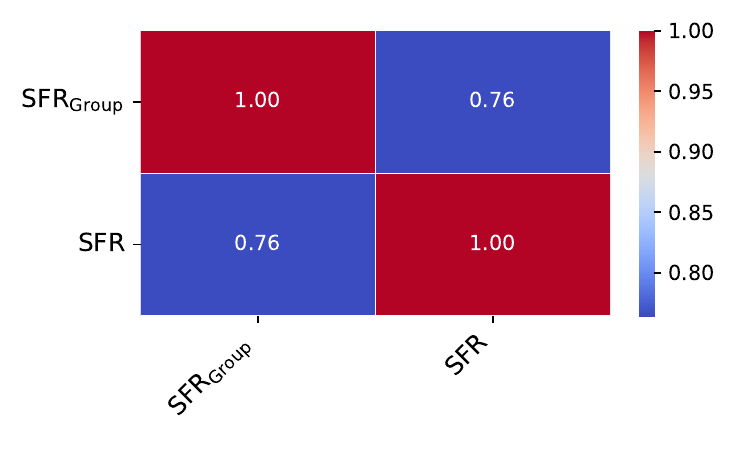}
    \caption{\textit{Correlation matrix for star formation rates for group and subhalo.}}
    \label{corr_SFR}
 \end{figure}

\end{document}